\documentclass[conference]{IEEEtran}

\usepackage{color}
\usepackage{verbatim, subfigure}
\usepackage{array, calc, multicol}
\usepackage{xspace}
\usepackage{graphics, epstopdf, graphicx, epsfig, psfrag, epsf, subfigure}
\usepackage{amssymb, amsfonts, amsmath, times}
\usepackage{multicol,balance, verbatim}

\usepackage{mathtools}

\DeclarePairedDelimiter\floor{\lfloor}{\rfloor}

\usepackage{algorithm}
\usepackage[noend]{algpseudocode}

\newcommand{\ie}{{\em i.e., }}

\newcommand{\eg}{{\em e.g., }}

\DeclareGraphicsExtensions{.eps, .pdf, .png, .jpg}   % optional

\newcommand{\Kset}{\mathcal{K}}

\newcommand{\Cset}{\mathcal{C}}

\newcommand{\Iset}{\mathcal{I}}

\newcommand{\MaCC}{{\tt PrComp}}

\IEEEoverridecommandlockouts

\begin{document}

\title{Predictive Edge Computing with Hard Deadlines}

\author{Yuxuan Xing, Hulya Seferoglu\\
{ECE Department, University of Illinois at Chicago}\\
{yxing7@uic.edu, hulya@uic.edu}
\thanks{This work was supported in part by U.S. Department of Commerce, National Institute of Standards and Technology award 70NANB17H188, and ARL grant W911NF-17-1-0032.}  \vspace{-15pt}
}

%\IEEEpubid{0000--0000/00\$00.00~\copyright~2012 IEEE}

\maketitle

%\IEEEpubid{978-1-5386-4533-8/18/\$31.00~\copyright~2018 IEEE}

\begin{abstract}

Edge computing is a promising approach for localized data processing for many edge applications and systems including Internet of Things (IoT), where computationally intensive tasks in IoT devices could be divided into sub-tasks and offloaded to other IoT devices, mobile devices, and / or servers at the edge. However, existing solutions on edge computing do not address the full range of challenges, specifically heterogeneity; edge devices are highly heterogeneous and dynamic in nature. In this paper, we develop a predictive edge computing framework with hard deadlines. Our algorithm; \MaCC \ (i) predicts the uncertain dynamics of resources of devices at the edge including energy, computing power, and mobility, and (ii) makes sub-task offloading decisions by taking into account the predicted available resources, as well as the hard deadline constraints of tasks. We evaluate \MaCC \ on a testbed consisting of real Android-based smartphones, and show that it significantly improves energy consumption of edge devices and task completion delay as compared to baselines. 

%Edge computing is a promising approach for localized data processing for Internet of Things (IoT), where computationally intensive tasks in IoT devices could be divided into sub-tasks and offloaded to other IoT devices, mobile devices, and / or servers at the edge. However, existing solutions on edge computing do not address the full range of challenges, specifically heterogeneity; edge devices are highly heterogeneous and dynamic in nature. In this paper, we develop a predictive edge computing framework with hard deadlines. Our algorithm; \MaCC \ (i) predicts the uncertain dynamics of resources of devices at the edge including energy, computing power, and mobility, and (ii) makes sub-task offloading decisions by taking into account (i) the predicted available resources, and (ii) task hard deadline constraints. We evaluate \MaCC \ on a testbed consisting of real Android-based smartphones, and show that it significantly improves energy consumption of edge devices and task completion delay as compared to baselines. 

\end{abstract}

\IEEEpubidadjcol
\vspace{-5pt}
\section{\label{sec:intro}Introduction}
\vspace{-5pt}
%The Internet of Things (IoT) is emerging as a new paradigm that connects an exponentially increasing number of devices, including smartphones, wireless sensors, smart meters, health monitoring devices, etc. 
The number of edge devices, \eg Internet of Things (IoT) keeps increasing and is estimated to reach billions in the next five years \cite{cisco_index}. As a result, the data collected by edge devices will grow at exponential rates. In many applications, unlocking the full power of edge devices requires analyzing and processing this data through computationally intensive algorithms with stringent latency constraints. 

In many scenarios, these algorithms cannot be run locally on the computationally-limited  edge devices (\eg IoT devices) and are rather outsourced to the cloud \cite{Kumar_computationa_off_survey}. Yet, offloading tasks to remote could be costly, inefficient in terms of delay, or may not be feasible (\eg when there is no  cellular or Wi-Fi infrastructure support). An alternative is edge computing, where tasks in an edge device could be offloaded to other edge devices including IoT devices, mobile devices, and / or servers in close proximity. %As illustrated in Fig.~\ref{fig:intro_example}, a master device can offload computationally intensive tasks to other devices (workers) in close proximity using device-to-device (D2D) connections. 

\begin{figure}[t!]
\centering
%\vspace{-5pt}
\includegraphics[height=25mm]{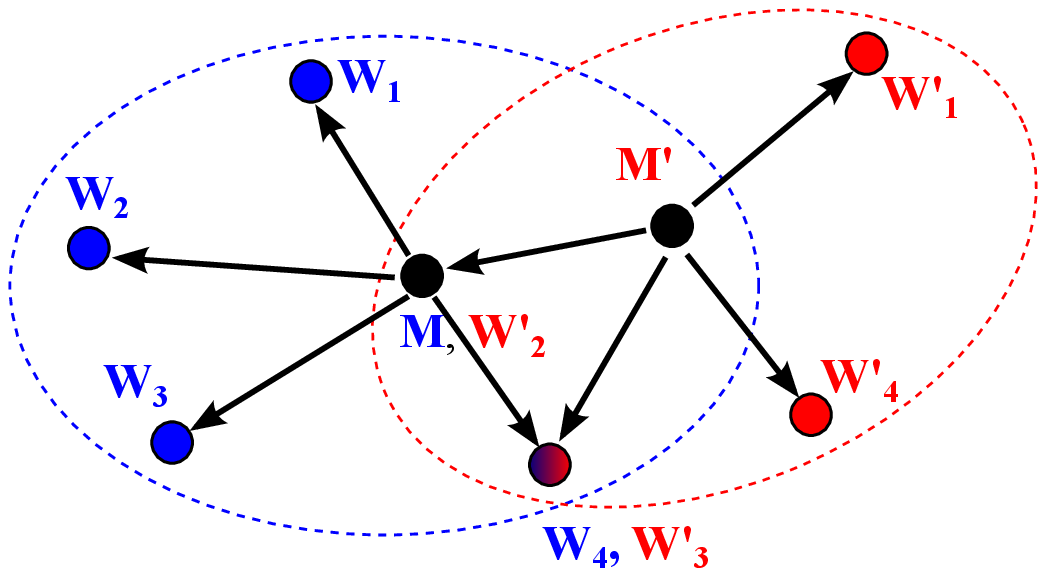}
\vspace{-5pt}
\caption{ Example computation at the edge. Two master devices $M$ and $M'$ offload their tasks to workers $W_1, \ldots, W_4$ and $W'_1, \ldots, W'_4$, respectively. As seen, a device could be (i) both a master and a worker at the same time, and (ii) workers of multiple masters simultaneously. 
}
\vspace{-20pt}
\label{fig:intro_example}
\end{figure}

However, existing solutions on edge computing do not address the full range of challenges, specifically heterogeneity; edge devices are highly heterogeneous and dynamic in nature. For example, if master device $M$ offloads some tasks to worker device $W_4$ in Fig.~\ref{fig:intro_example}, but if device $W_4$ is running another computationally intensive application (either originated from itself or offloaded from another master device - $M'$ in this example), delay increases. Similarly, if device $M$ offloads some tasks to device $W_4$, but before completing processing these tasks, device $W_4$ moves away, D2D connection between $M$ and $W_4$ is broken. In this case, device $M$ should offload its tasks again to other devices in close proximity (\eg device $W_3$). This re-offloading process increases delay and energy consumption. Thus, we should develop an edge computing mechanism, which is aware of the heterogeneity and time varying nature of resources as well as mobility of devices. 

In this paper, we develop a predictive computation offloading mechanism (\MaCC) by taking into account heterogeneous and time varying resources as well as mobility. In particular, we consider a master / worker setup as seen in Fig.~\ref{fig:intro_example}, where a master device predicts the resources of workers using periodic probes. Then, the master device makes computation offloading decisions to minimize the energy consumption of master and worker devices while satisfying deadline constraints of tasks.

We evaluate \MaCC \ on a testbed consisting of real smartphones, and show that it significantly improves energy consumption and task completion time as compared to baselines. The following are the key contributions of this work:
\begin{itemize}
\vspace{-2pt}
\item  We develop a resource prediction module for Android-based mobile devices. Our prediction module determines the amount of delay and energy consumption when a task is processed locally or remotely in an Android device. This module also predicts the mobility of devices. 
\vspace{-1pt}
\item We develop online task scheduling algorithms \MaCC \ for serial and parallel tasks by using the predicted available resources. Our algorithms are based on the structure of the solution of an optimal task scheduling problem. 
\vspace{-1pt}
\item We evaluate \MaCC \ on a testbed consisting of real smartphones, and show that it brings significant performance benefits in terms of energy consumption and delay. 
\vspace{-5pt}
%\textcolor{red}{Prediction. We formulate an optimal mobility-aware cooperative computation problem. The solution of the problem gives an offline computation algorithm that takes into account the mobility of devices as well as deadline of tasks and energy of devices.}
%\item \textcolor{red}{Algorithm design. We develop an online algorithm based on the structure of the offline algorithm. The online algorithm makes offloading decisions on the fly efficiently by still taking into account mobility of devices.}
%\item \textcolor{red}{Evaluation.We evaluate our \MaCC \ framework on a testbed consisting of real smartphones, and we show that it brings significant performance benefits in terms of reliability, delay, and energy consumption. }
\end{itemize}

%~~

%~~

The structure of the rest of the paper is as follows. Section~\ref{sec_related} presents related work. Section~\ref{sec:system} gives an overview of the system model. Section~\ref{sec:res_prediction} presents our delay, energy, and mobility prediction module. Section~\ref{sec:algorithms} presents our  \MaCC \ algorithms. Section~\ref{sec:results} evaluates the performance of our algorithms in a real testbed. Section~\ref{sec:conclusion} concludes the paper.

\IEEEpubidadjcol
%\vspace{-5pt}
\section{\label{sec_related} Related Work}
%\vspace{-5pt}
Mobile cloud computing is a rapidly growing field with the goal of providing extensive computational resources to mobile devices as well as higher quality of experience \cite{survey_mobile_cloud1}, \cite{survey_mobile_cloud2}, \cite{survey_mobile_cloud3}. 
%, \cite{survey_mobile_cloud4}, \cite{survey_mobile_cloud5}. 
The initial approach to mobile cloud computing has been to offload resource intensive tasks to remote clouds by exploiting Internet connectivity of mobile devices. This approach has received a lot of attention which led to extensive literature in the area \cite{comet}, \cite{maui}, \cite{DPartner}, \cite{Cuckoo}, \cite{MCC_resource_allocation1}. The feasibility of computation offloading to remote cloud by mobile devices \cite{thinkair_dynmic_resource_allocation_and_parallel_execution_in_cloud_for_mobile_code_offloading} as well as energy efficient computation offloading \cite{energy_efficient_computation_offloading_in_cellular_networks, energy_efficient_scheduling_policy_for_collaborative_exec_in_mobile_cloud_computing} has been considered in the previous work. As compared to this line of work, our focus is on edge computing rather than remote clouds.

There is an increasing interest in edge computing by exploiting connectivity among mobile devices \cite{survey_mobile_to_ubiq}. This approach suggests that if devices in close proximity are capable of processing tasks cooperatively, then local area computation groups could be formed and exploited for computation. This approach, sparking a lot of interest, led to some very interesting work in the area \cite{trans_clouds1}, \cite{cloud_down_to_earth}, \cite{mclouds}. The performance of computing at the edge including the computation group size of IoT devices, lifetime, and reachable time is characterized in \cite{can_mobile_cloudlets_support_mobile_apps} by taking into account the mobility of devices. As compared to this line of work, we focus on offloading tasks from one edge device to others by predicting delay, energy, and mobility. 

Edge computing is investigated in \cite{on_the_computation_offloading_at_ad_hoc_cloudlet} to deal with mobility by exploiting both cellular and Wi-Fi connections. 
As compared to this work, mobile devices can communicate only when they are in close proximity in our work, and we make offloading decisions by taking into account the mobility patterns of devices. 
Task offloading to minimize energy consumption of  master devices is considered in \cite{ARC}. As compared to \cite{ARC}, our goal is to minimize the energy consumption of both master and worker devices by taking into account hard deadlines.

%\IEEEpubidadjcol
%\vspace{-5pt}
\section{\label{sec:system} Setup and Problem Statement}
%\vspace{-5pt}
{\em Topology.} We consider a setup illustrated in Fig.~\ref{fig:intro_example} with multiple master and worker devices.  We particularly focus on a master / worker cluster to make presentation simple. We assume that there is a {\em master} device and $N$ {\em worker} devices in a cluster. We define a set $\Cset = \{M, W_1,  \ldots W_N\}$ as the set of all devices in a cluster. 
The device $M \in \Cset$ is the {\em master} device that creates tasks and offloads them to other devices. The set $\{ W_1, W_2 \ldots W_N\} \in \Cset$ represents {\em worker} devices that are in close proximity of the master device. 

%\begin{figure}[t!]
%\centering
%{ {\includegraphics[height=42mm]{int_figs/setup_fig_v2.eps}} } 
%\vspace{-5pt}
%\caption{\textcolor{red}{REVISE THIS FIGURE. System setup. $\eta_0$ is the device that offloads tasks to other mobile devices, servers, and the remote cloud in the set $\{ \eta_1, \eta_2 \ldots \eta_N\}$.}
%}
%\vspace{-10pt}
%\label{fig:setup}
%\end{figure}

%In our setup, we consider that a device can either offload its tasks, or receive tasks from another device to process. For example, only $\eta_0$ can offload tasks to other devices in Fig.~\ref{fig:setup}, while the devices in the set $\{ \eta_1, \eta_2 \ldots \eta_N\} $ can only process tasks. \textcolor{red}{It is trivial to extend our work to the case that each device both serves as an originator and helper}.  

{\em Offloading Scenario \& Applications.} We consider a scenario that a master device $M$ runs multiple applications, and offloads computationally intensive applications to worker devices. %An example scenario could be the case that  $M$ streams a video content from a remote server (\eg from Youtube). Simultaneously, $M$ can run computationally intensive applications such as searching a text file to find a specific string, virus scanning, face detection, solving N-queens puzzle, matrix multiplication \cite{thinkair_dynmic_resource_allocation_and_parallel_execution_in_cloud_for_mobile_code_offloading}.  
In this paper, we focus on face detection and matrix multiplication applications. % in our implementation using Android-based smartphones.

Face detection application processes a number of images, and detects human faces in each image. We employ Android SDK's FaceDetector and FaceDetector.Face classes to implement a face detection application, which puts a circle on the faces it detects. % as demonstrated in Fig.~\ref{fig:face_detection_example}. 
The processed image is stored in the external memories of devices. 

%@MISC{face_detection_software,
%   title = "Captured video of MaCC - Smooth playout",
%   url = "http://www.anddev.org/quick_and_easy_facedetector_demo-t3856.html",
%}

%from a image that is stored on the external memory of a mobile device. The detection procedure is implemented by series of built-in Android graph and media APIs. After detection, a dot is put between eyes of every detected face. Finally, the output image will be stored on the external memory. Fig. \ref{face} demonstrates the original image and processed image.
%(http://www.anddev.org/quick_and_easy_facedetector_demo-t3856.html)Note that not every face will be recognized.

%\begin{figure}[t!]
%\centering
%%\vspace{-2pt}
%\subfigure[Original Image]{ {\includegraphics[height=28mm]{Fig_Yuxuan/manyfaces.pdf}} }
%\subfigure[Processed Image]{ {\includegraphics[height=28mm]{Fig_Yuxuan/faceDetected.pdf}} }
%\vspace{-5pt}
%\caption{Face detection example using Android SDK's FaceDetector and FaceDetector.Face classes. 
%}
%\vspace{-10pt}
%\label{fig:face_detection_example}
%\end{figure}

Matrix multiplication application computes $Y=AX$ where $A = (a_{i,j}) \in \mathbb{R}^{R_1 \times R_2}$, $X = (x_{i,j}) \in \mathbb{R}^{R_2 \times R_3}$, and $Y = (y_{i,j}) \in \mathbb{R}^{R_1 \times R_3}$. Our application uses simple matrix multiplication, \ie computes $y_{i,j} =  \sum_{r = 1}^{R_2} a_{i,r} x_{r,j}$, $\forall i,j$.

%A of size $m$ by $n$ and B of size $n$ by $l$, matrix multiplication is to produce the product of A and B, which is a $m$ by $l$ matrix. In our experiments, we use the naive approach to calculate the multiplication. To be precise:
%$$(AB)_{i,j} = \sum_{k=1}^{n}A_{i,k}B_{k,j}$$
%where $(AB)_{i,j}$ is the element on $i$th row and $k$th column of the output matrix. The same operation is performed for all $i=1,2,...m$ and $j=1,2,...l$.  

%In this setup, computationally intensive applications put a strain on computing power and memory, which hurts the video streaming application by introducing freezing scenes. To support this idea in practice, we conducted a pilot study, where a mobile device streams video from Youtube via high-speed Wi-Fi link (\ie bandwidth is not a bottleneck). Meanwhile, the mobile device runs a text searching algorithm, which tries to find a particular string in a large text file. The details of the search algorithm as well as experiment conditions will be provided in Section~\ref{sec:performance}. In this experiment, we have observed that the computationally intensive text searching task introduces freezing scenes in video. In particular, 20\% of the time, we have observed freezing scenes. This small pilot study shows that it is critical to postpone computationally intensive tasks so that they do not hurt the quality of other applications. However, if the tasks are delay constrained, then the only solution is computation offloading. 

{\em Task Model.} We consider a set of tasks; $\Kset = \{ \mbox{Task }1, \ldots, \mbox{Task } K \}$. We consider two types of tasks: (i) serial, and (ii) parallel. When the tasks are serial, $k$th task can only be processed after the $k-1$th task is processed. Parallel tasks could be processed simultaneously at multiple workers. 

%\begin{figure}[t!]
%\vspace{5pt}
%\centering
%\subfigure[Serial tasks]{ {\includegraphics[width=85mm]{int_figs/task_model.eps}} } 
%\subfigure[\textcolor{red}{HS: Revise. Parallel tasks}]{ {\includegraphics[width=85mm]{int_figs/task_model.eps}} } 
%\vspace{-5pt}
%\caption{(a) Serial task model, where Task $k$ can processed only after Task $k-1$ is processed. (b Parallel task model, where all the tasks can be processed simultaneously.)}
%\vspace{-10pt}
%\label{fig:task_model}
%\end{figure}

%The size of Task $k$ is $B_k$ bytes. The workload of Task $k$ is $\omega_k$, and could be different than its original size $B_k$. Indeed, a task could be computationally intensive even its original size is small. Thus, workload $\omega_k$ reflects this variation. Furthermore, the size of Task $k$ could vary (increase or decrease) after processing, and $\alpha_k > 0$ shows this variation. If $\alpha_k < 1$, the size of Task $k$ decreases after processing. Otherwise, \ie $\alpha_k > 1$, the size of Task $k$ increases after processing. 

{\em Offloading Policy.} We assign each task to a device in the set $\Cset$. However, due to mobility, a master or a worker device may move after a task is offloaded, so they may be out of transmission range of each other. Thus, task offloading becomes unsuccessful (worker cannot send the processed task back to the master). In this case, the task should be rescheduled again. Assume that $t_{k,l}$ is the time that Task $k$ is scheduled for the $l$th time, and $\pi_{k,l}$ is the policy that shows at which device that Task $k$ is scheduled at the $l$th trial. Thus, $\pi_{k,l} \in \Cset$. For example, if the $k$th task is scheduled to be processed at the master device $M$ (\ie if the task is not offloaded to any worker) at the $l$th trial, then $\pi_{k,l} = M$. 

Let us assume that the set of policies $\boldsymbol \pi_k = \{\pi_{k,1}, \ldots, \pi_{k,L_k}\}$ corresponds to the policy for Task $k$, where $L_k$ is the last scheduling trial of Task $k$. Note that $L_k$ depends on the optimal policy as well as the randomness due to mobility. The set $ \boldsymbol \pi = \{ \boldsymbol \pi_1,  \ldots,  \boldsymbol \pi_K  \}$  corresponds to the policy for scheduling all the tasks. 

{\em Problem Statement.} Our goal is to determine a policy $ \boldsymbol \pi$ that minimizes the  total energy consumption at all devices (master and workers) subject to hard deadline constraints by estimating per-task energy consumption and delay at the master and worker devices as well as predicting the mobility of workers.

%\vspace{-5pt}
\section{\label{sec:res_prediction}  Delay, Energy, and Mobility Prediction}
%\vspace{-5pt}
In this section, we present our approach for predicting energy, delay, and mobility of master and worker devices using Android-based mobile devices. The results of this section will be used in our algorithm design in the next section. 

We implemented a testbed of a master and multiple worker cluster using real mobile devices, specifically Android 6.0.1 based Nexus 6P smartphones. All the workers are connected to the master device using Wi-Fi Direct connections in our testbed. (In other words, the master device is configured as the group owner of the Wi-Fi Direct group, which is a star topology.) We conducted our experiments using our testbed in a realistic lab environment, where several other Wi-Fi networks were operating in the background. We located all the devices in close proximity of each other  (within a few meters distance). We use face detection application as a demonstrating example for our prediction. 

\vspace{-5pt}
\subsection{\label{sec:delay} Delay} 
\vspace{-5pt}
We determine task delay as the amount of time for offloading, processing, and receiving tasks back from the worker devices. Each master device in our predictive edge computing framework determines per task delay by periodically probing itself as well as its workers. In particular, the master device puts a timestamp to each task before processing or offloading it to a worker. For example, the time stamp of the $i$th task such that $i=kl$ is $t_i$.\footnote{Note that if device (master or worker) is not assigned any tasks due to our scheduling algorithm, we still offload tasks to this device periodically to predict its resources.}.

%the $i$th task trial such that $i=kl$, the time stamp becomes $t_i$. 

Assume that $\tilde{t}_{i,0}$ is the time that the master devices completes the task by processing locally (no offloading), and $\tilde{t}_{i,n}$ is the time that receives the completed task from worker $W_n$. Let us define $\Iset_{\Cset} = \{0, 1, \ldots, N\}$ as an index set of $\Cset$, where $0 \in \Iset_{\Cset}$ corresponds to the master device, and $n \in \Iset_{\Cset}$ corresponds to worker $W_n$. Thus, the delay for each device $j \in \Iset_{\Cset}$ becomes $\theta_{i,j} =  \tilde{t}_{i,j} -  t_{i}$. Note that $\theta_{i,j}$ includes only processing delay at the master device when $j = 0$, and it includes offloading and processing times as well as the time to receive tasks back from the worker devices when $j \neq 0$. 

\begin{figure}[t!]
\centering
\vspace{-10pt}
%\includegraphics[height=40mm]{Fig_Yuxuan/basic_time_new.eps}
%\subfigure[Old Nexus 6P]{ {\includegraphics[height=29mm]{Fig_Yuxuan/basic_time_old.eps}} }
%\subfigure[New Nexus 6P]{ {\includegraphics[height=29mm]{Fig_Yuxuan/basic_time_new.eps}} }
{\includegraphics[height=45mm]{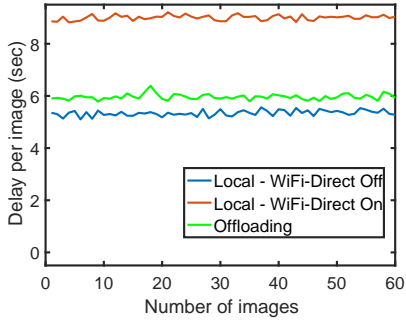}} 
\vspace{-5pt}
\caption{Average delay per image for a master / worker setup. %{\tt Local - Wi-Fi Direct On} and {\tt Local - Wi-Fi Direct Off} correspond to the scenarios that master device locally processing the images when (i) it is connected to another device via Wi-Fi Direct and (ii) Wi-Fi Direct connection is closed, respectively. {\tt Offloading} is the scenario that the master device offloads the images to its worker.
}
\vspace{-20pt}
\label{fig:proc_time}
\end{figure}

Fig.~\ref{fig:proc_time} presents delay versus number of images for a master / worker setup. %Both master and worker devices are Nexus 6P smartphones. We consider face detection application, where each image and detecting faces in it corresponds to a task. At a given time master either locally processing tasks or offloading them to the worker device. 
The amount of delay  is the average of 10 trials. %Fig.~\ref{fig:proc_time}(a) shows the results when the first release of Nexus 6P smartphones are used, while Fig.~\ref{fig:proc_time}(a) shows the results with the second release of Nexus 6P smartphones. 
{\tt Local - Wi-Fi Direct On} and {\tt Local - Wi-Fi Direct Off} correspond to the scenarios that master device locally processing the images when (i) it is connected to another device via Wi-Fi Direct and (ii) Wi-Fi Direct connection is closed, respectively. {\tt Offloading} is the scenario that the master device offloads the images to its worker. As seen, the delay of {\tt Local - Wi-Fi Direct Off} is less than both {\tt Local - Wi-Fi Direct On} and {\tt Offloading}, because it locally processes the packet, and no time is wasted for transmitting packets. On the other hand, {\tt Offloading} is better than {\tt Local - Wi-Fi Direct On}, because when a master device opens Wi-Fi Direct connection on and becomes a group owner (\ie behaves as an access point), it has more computational load, which increases delay. This figure shows that (i) delay characteristics of devices can be measured by probing these devices, and (ii) delay performance of a device is heterogeneous and can be time-varying depending on its configuration. %, \eg if it is a group owner of a Wi-Fi Direct group or not. 

\vspace{-5pt}
\subsection{\label{sec:energy}Energy}
\vspace{-5pt}
The main source of energy consumption in edge computing applications comes from computing and offloading tasks. This section deals with predicting energy consumption due to CPU usage and packet transmission (and reception) using Wi-Fi interfaces in Android-based devices. However, Android APIs do not provide granular (per application and hardware) energy consumption. 
%only provide total energy consumption per application or per CPU or Wi-Fi energy consumption across all applications, so it is challenging to determine per application and per CPU / Wi-Fi energy consumption. 
%Previous work \textcolor{red}{[REF REF]} uses auxiliary devices to measure energy consumption. 
Next, we present our approach to predict energy consumption without using any external devices. 

{\em Energy consumption due to computation (\ie CPU).}  
The modern CPUs of Android devices consist of multiple clusters, and each cluster can operate at different speeds. Let $\iota_{c,s}$ is the amount of electrical current (in mA) that cluster $c$ uses when operating at speed $s$, which is not time-varying, and can be found on power profile of every Android device. 

%The modern CPUs of Android devices consist of multiple clusters \textcolor{red}{HS: How many?}, and each cluster can operate at different speeds \textcolor{red}{HS: How many different speeds? An example}. Let $\iota_{c,s}$ is the amount of electrical current (in mA) that cluster $c$ uses when operating at speed $s$, which is not time-varying, and can be found on power profile of every Android device. 

When a computationally intensive application is run on a device, multiple clusters at different speeds could be used. If we can predict the amount of time that each cluster - speed pair is used, we can characterize the amount of battery power (in mAh) used per application. Although Android APK does not provide this information directly, the following information can be gathered. 

Let (i) $T_a(t)$ be the total amount of time that application $a$ has used CPU (across all clusters and speeds) since the device is plugged off from a power supply (let us denote this time $t_0$) until time $t$, and (ii) $\tau_{c,s}(t)$ be the amount of time cluster $c$ is used at speed $s$ between $t_0$ and $t$. Both $T_a(t)$ and $\tau_{c,s}(t)$ information can be acquired using the class \textit{BatteryStatsHelper} of Android in the form of a list of ``battery sippers''. 

Each battery sipper represents an application associated with a unique application ID. The application ID of a desired application can be found in the \textit{Process} Android class. The battery sipper has the battery related information including both $T_a(t)$ and $\tau_{c,s}(t)$. Next, we define per application battery consumption using these parameters. 

The battery consumption due to application $a$ between time interval $t-\delta$ and $t$, where $\delta$ is a small time interval, is expressed as
\begin{align}\label{eq:energy}
& e_a^{\text{CPU}}(t, t-\delta) = \frac{T_a(t) - T_a(t-\delta)}{T_{\text{all}}(t) - T_{\text{all}}(t-\delta)} \sum_{\forall c} \sum_{\forall s} (\tau_{c,s}(t) - \nonumber \\
& \tau_{c,s}(t-\delta)) \iota_{c,s}, 
\end{align} where $T_{\text{all}}(t)$ is the total amount of time that CPU is used for all applications. Although $T_{\text{all}}(t)$ cannot be directly gathered from Android APK, we can characterize it as $T_{\text{all}}(t) = \sum_{\forall c} \sum_{\forall s} \tau_{c,s}(t)$.

Note that the term $\frac{T_a(t) - T_a(t-\delta)}{T_{\text{all}}(t) - T_{\text{all}}(t-\delta)}$ in Eq.~(\ref{eq:energy}) represents the percentage of time that application $a$ uses available CPU resources as compared to all other applications. On the other hand, $\sum_{\forall c} \sum_{\forall s} (\tau_{c,s}(t) -  \tau_{c,s}(t-\delta)) \iota_{c,s}$ represents the total energy consumption between time $t-\delta$ and $t$ for all applications. The multiplication of these two terms is a good predictor of the energy consumption by application $a$ between time interval $t-\delta$ and $t$. 

\begin{figure}[t!]
\centering
\vspace{-10pt}
\includegraphics[height=40mm]{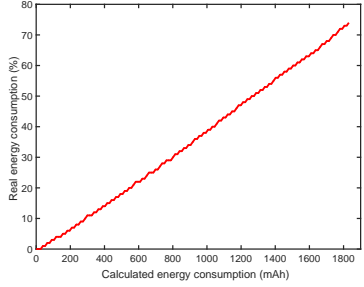}
\vspace{-5pt}
\caption{Real versus calculated energy consumption for face detection application at the master device.}
\vspace{-15pt}
\label{fig:energy_sanity_check}
\end{figure}

Now, let us assume that $i$th task of application $a$ is processed at device $j$ at time $t - \delta$, and the total processing time is $\delta$. Thus, we can characterize the amount of battery consumption due to CPU for processing task $i$ at device $j$ as $\epsilon_{i,j}^{\text{CPU}} = e_a^{\text{CPU}}(t,t-\delta)$. Fig.~\ref{fig:energy_sanity_check} shows the real and calculated energy consumption for face detection application at the master device. In particular, the master device  (Nexus 6P) detects faces in multiple images one by one. There is no other user-level applications running on the device, and it operates in the {\em airplane mode}, so all network interfaces are closed (\ie no other energy consumption). The x-axis shows the cumulative energy consumption, \ie $\sum_{\forall \alpha}^{i} \epsilon_{\alpha,j}^{\text{CPU}}$, while the y-axis shows the battery drop percentage directly read from the device. Considering that the battery capacity of Nexus 6P devices is 3450 mAh, our energy calculation is a good predictor of per application energy consumption. 

{\em Energy consumption due to Wi-Fi.} We measure this type of energy consumption using the energy consumption profile of Wi-Fi interface. (Note that Wi-Fi and Wi-Fi Direct share the same interface.) Unlike CPU energy consumption, it is straightforward to measure energy consumption in Android for using Wi-Fi interface. In particular, \textit{wifiPowerMah} value is stored in the battery sipper discussed above, and it represents the total energy consumption to keep Wi-Fi interface open, transmitting, and receiving packets. Thus, we can directly obtain the battery consumption $e_a^{\text{Wi-Fi}}(t, t-\delta)$ due to application $a$ between time interval $t-\delta$ and $t$, where $\delta$ is a small time interval. If $i$th task from application $a$ is processed at device $j$ during $t-\delta$ and $t$, the battery  consumption due to Wi-Fi interface becomes $\epsilon_{i,j}^{\text{Wi-Fi}} = e_a^{\text{Wi-Fi}}(t, t-\delta)$. 

%After the CPU energy consumption is known, the next interest would be how to measure Wi-Fi energy consumption for a particular application. Unlike CPU, Wi-Fi energy consumption can be directly obtained by the information (\textit{wifiPowerMah}) stored in the battery sipper discussed above. The Wi-Fi energy consumption for the originator and helper $n$ is denoted by $e_{wifi}^o$ and $e_{wifi}^n$ respectively.
 
\begin{figure}[t!]
\centering
%\vspace{-10pt}
%\includegraphics[height=40mm]{Fig_Yuxuan/basic_time_new.eps}
\subfigure[CPU]{ {\includegraphics[height=33mm]{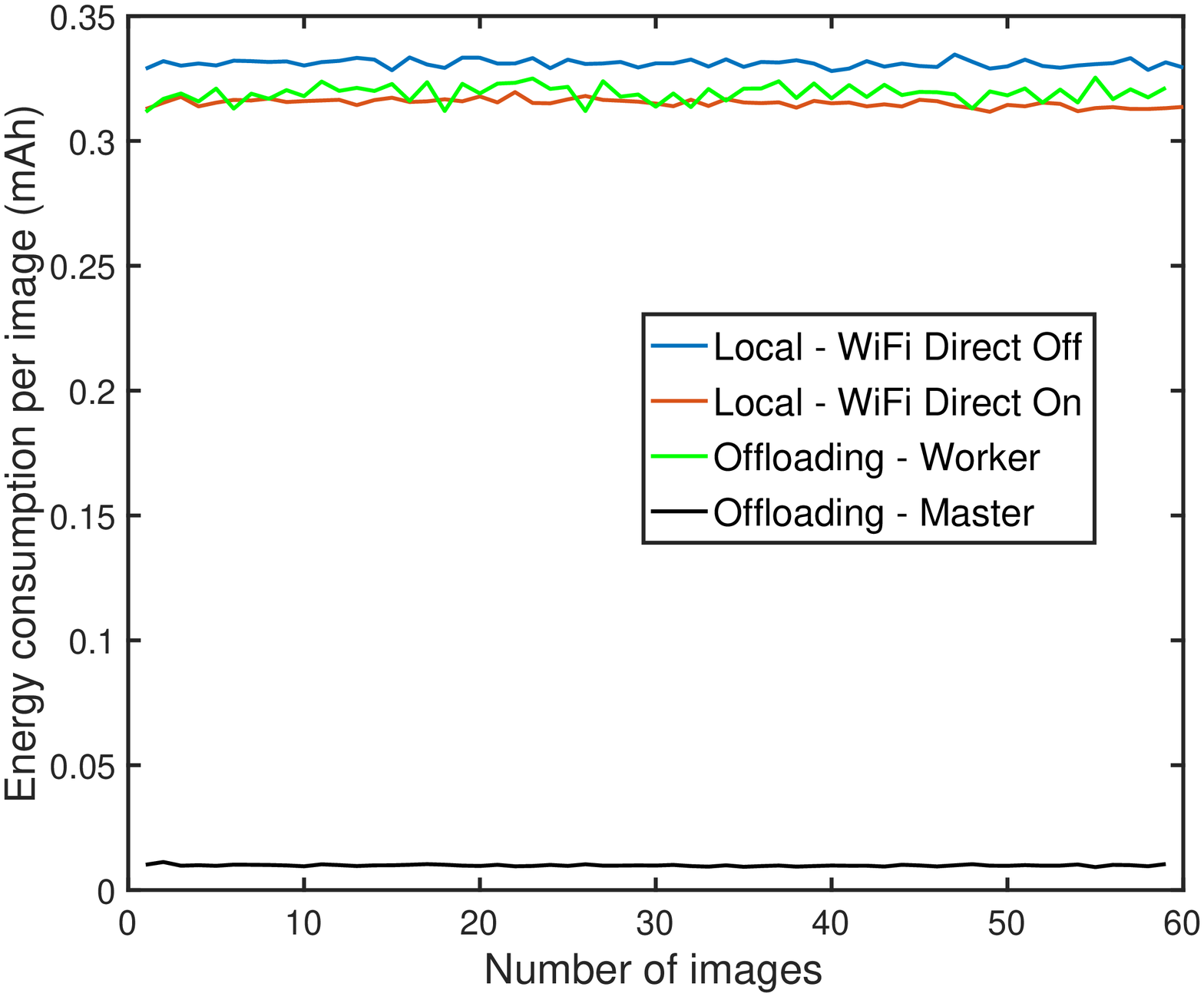}} } \hspace{-20pt}
\subfigure[Wi-Fi]{ {\includegraphics[height=32mm]{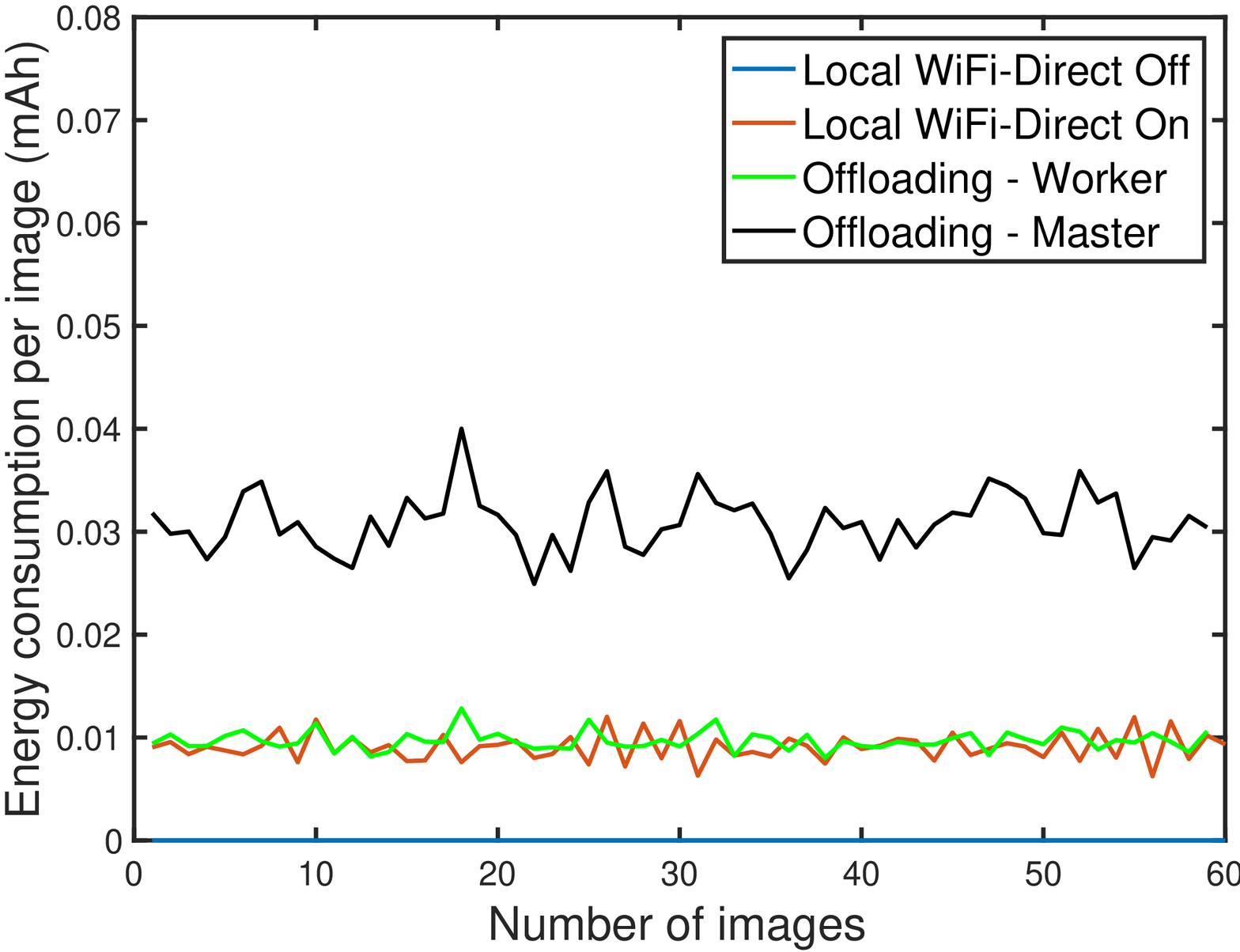}} }
\vspace{-5pt}
\caption{Average energy consumption due to (a) computation (\ie CPU), (b) transmission and reception (\ie Wi-Fi).}
\vspace{-15pt}
\label{fig:proc_energy}
\end{figure}

The energy consumption per image due to CPU and Wi-Fi is presented in Fig.~\ref{fig:proc_energy}. Each graph is an average of 10 experiments. {\tt Offloading - Master} and {\tt Offloading - Worker} are the energy consumption at master and worker devices when tasks are offloaded to a worker device. {\tt Local - Wi-Fi Direct On} and {\tt Local - Wi-Fi Direct Off} are the same as in Fig.~\ref{fig:proc_time}. 
As seen, energy consumption due to CPU is higher than Wi-Fi and changes depending on whether a device is processing or offloading a task. %\textcolor{red}{Talk about legends.} 

%The average energy consumption per image for CPU and Wi-Fi can be found in Fig. \ref{basic_cpu} and Fig. \ref{basic_wifi}. There are 10 trials in each plot. 
%The interesting point is that the average Wi-Fi energy consumption for the old devices is slightly higher than the new ones. As the results shown in Fig. \ref{basic_time}, offloading in old devices requires longer time to complete. Therefore, old devices demand Wi-Fi to be on for a longer time thus consume additional energy.

%The hardware of the old and new Nexus 6P devices are exactly the same and they both run on Android 6.0.1. The only difference is their Linux kernal version, baseband version and build number.

\vspace{-5pt}
\subsection{\label{sec:mobility} Mobility}
\vspace{-5pt}
We consider three types of mobility models: (i) {\em Statistical:} The probability that a master and worker devices are in the same transmission range is known a priori. (ii) {\em Predicted:} A master device predicts that a worker device is in its transmission range with some error margin. (iii) {\em Majority voting:} A master device uses history to predict the mobility of itself and workers. In particular, a master device divides the time into slots and checks the most recent encounters (\ie being in the same transmission range) with a worker. If during most of the recent slots, there is encounter with a master and worker, the master decides that they will be in the same transmission range in the next slot.

\section{\label{sec:algorithms} \MaCC \ Algorithms}
%\vspace{-5pt}
In this section, we develop \MaCC \ algorithms for serial and parallel tasks. Our  \MaCC \ algorithms are based on the solution to the optimal task allocation, and use the predicted delay, energy consumption, and mobility in Section~\ref{sec:res_prediction}. 

\vspace{-5pt}
\subsection{Serial Tasks}
\vspace{-5pt}
%In this section, we consider serial tasks. 
%Let us assume that $k-1$ tasks are already scheduled, and the $k$th task will be scheduled for the $l$th time. Our goal is to determine the policy $\pi_{k,l}$, \ie the device  at which the $l$th trial of $k$th task is scheduled. 
Our first step is to solve the following optimization problem.  
\begin{align}\label{eq:serial_opt_problem}
\text{min}_{\{k_n\}_{n \in  \Iset_{\Cset} }} & \sum_{n \in \Iset_{\Cset}}  E(n) k_n \nonumber \\
\text{subject to} &  \sum_{n \in \Iset_{\Cset}} \Delta(n) k_n \leq \Delta_{\text{thr}} \nonumber \\
&  \sum_{n \in \Iset_{\Cset}} k_n = K, 
\end{align} where $E(n)$ and $\Delta(n)$ are the average energy consumption and delay for processing one task at device $n$, $k_n$ is the number of tasks assigned to device $n$, and $\Delta_{\text{thr}}$ is the hard deadline constraint for processing tasks. (We will describe how $E(n)$ and $\Delta(n)$ are calculated later in this section.)  The objective function of (\ref{eq:serial_opt_problem}) is to minimize the total energy consumption at master and worker devices. The first constraint is the deadline constraint, and the last constraint makes sure that all $K$ tasks are scheduled. 

The optimal solution to (\ref{eq:serial_opt_problem}) selects $n^{*} = \arg \min E(n)$ that satisfies $\Delta(n^*)K \leq \Delta_{\text{thr}}$, and allocates all tasks to $n^*$, \ie $k_{n^*} = K$. Note that this is an offline solution that makes task offloading decisions prior to scheduling. However, in heterogeneous and time-varying systems, an online solution that makes a decision for each task is better as it is adaptive.

Our online \MaCC \ algorithm is based on the offline solution, and determines device $n^* = \pi_{k,l}^*$ at time $t_{k,l}$ for the $l$th trial of the $k$th task according to the following rule: Determine policy $\pi_{k,l}^* = \arg \min E(\pi_{k,l})$ that satisfies $\Delta(\pi_{k,l}) (K -k + 1) \leq \Delta_{\text{thr}} - t_{k,l}$. 
Next, we describe how to determine $\Delta(\pi_{k,l})$ and $E(\pi_{k,l})$ using our predicted values in Section~\ref{sec:res_prediction}. 

The average delay $\Delta(\pi_{k,l})$ at device $n = \pi_{k,l}$ for the $k$th task at $l$th trial is expressed by taking into account mobility as 
\begin{align} \label{eq:delay}
\Delta(\pi_{k,l}) =  \Big( \sum_{i=1}^{kl-1}  \frac{\theta_{i,\pi_{k,l}}}{kl-1} \Big) \frac{1}{1 - P_{\pi_{k,l}}}, 
\end{align} 
where $\theta_{i,\pi_{k,l}}$ is the delay that is measured as described in Section~\ref{sec:delay} and $P_{\pi_{k,l}}$ is the probability that device $\pi_{k,l}$ will not be in the transmission range of the master device. In this formulation, $ \sum_{i=1}^{kl-1}  \frac{\theta_{i,\pi_{k,l}}}{kl-1} $ is the average delay of all per-task delays until $i=kl$th task, and $\frac{1}{1 - P_{\pi_{k,l}}}$ reflects the contribution of the mobility on average delay. 

The average energy consumption is formulated as
\begin{align}
E(\pi_{k,l}) = \frac{\tilde{\epsilon}_{kl-1,\pi_{k,l}}^{\text{Proc}} + \tilde{\epsilon}_{kl-1,\pi_{k,l}}^{\text{Off}} }{1 - P_{\pi_{k,l}}}
\end{align} where $\tilde{\epsilon}_{kl-1,\pi_{k,l}}^{\text{Proc}}$ and $\tilde{\epsilon}_{kl-1,\pi_{k,l}}^{\text{Off}}$ are processing and offloading energy consumptions, respectively. The processing energy consumption at device $j$ for the $i=kl$th task is formulated as 
\begin{align}\label{eq:ep_proc}
& \tilde{\epsilon}_{i+1,j}^{\text{Proc}} = \Big( \tilde{\epsilon}_{i,j}^{\text{Proc}} \beta + (\epsilon_{i,j}^{\text{CPU}} + \epsilon_{i,j}^{\text{Wi-Fi}} 1_{[j \neq 0]} )\tilde{\beta}  \Big) 1_{[i \rightarrow j]} +  \tilde{\epsilon}_{i,j}^{\text{Proc}} \nonumber \\
& (1 - 1_{[i \rightarrow j]}) 
\end{align} where $\beta$ is a small constant, $\tilde{\beta} = 1 - \beta$,  $1_{[x]}$ is an indicator function and takes value $1$ if $x $ is true, and $0$ otherwise. $i \rightarrow j$ represents (a mapping) that task $i$ is offloaded to device $j$. Note that $\epsilon_{i,j}^{\text{CPU}}$ and $\epsilon_{i,j}^{\text{Wi-Fi}}$ are measured as described in Section~\ref{sec:energy}.  The term $(\epsilon_{i,j}^{\text{CPU}} + \epsilon_{i,j}^{\text{Wi-Fi}} 1_{[j \neq 0]} )$  in (\ref{eq:ep_proc}) states that there is always energy consumption due to CPU, but there is energy consumption due to Wi-Fi only when the task is offloaded from the master device to worker devices (\ie when $j \neq 0$). $\tilde{\epsilon}_{i,j}^{\text{Proc}} (1 - 1_{[i \rightarrow j]})$ term shows that processing energy consumption is updated only if task $i$ is offloaded to device $j$, \ie when $i \rightarrow j$ mapping is true. Similarly, the energy consumption due to offloading is expressed as  
\begin{align}\label{eq:ep_off}
\tilde{\epsilon}_{i+1,j}^{\text{Off}} = \Big( ( \tilde{\epsilon}_{i,j}^{\text{Off}} \beta + \epsilon_{i,j}^{\text{Wi-Fi}} \tilde{\beta} ) (1 - 1_{[i \rightarrow j]}) + \tilde{\epsilon}_{i,j}^{\text{Off}} 1_{[i \rightarrow j]} \Big) 1_{[j = 0]}
\end{align} where $\tilde{\epsilon}_{i+1,j}^{\text{Off}} = 0$ when $j \neq 0$, because only the master device (\ie when $j = 0$) offloads tasks to worker devices. Note that both (\ref{eq:ep_proc}) and (\ref{eq:ep_off}) assumes that $\tilde{\epsilon}_{0,j}^{\text{Proc}} = 0$, $\tilde{\epsilon}_{0,j}^{\text{Off}} = 0$, $\forall j$. 

\vspace{-5pt}
\subsection{Parallel Tasks}
\vspace{-5pt}
Our first step is to solve the following optimization problem
\begin{align}\label{eq:parallel_opt_problem}
\text{min}_{\{k_n\}_{n \in  \Iset_{\Cset} }} & \sum_{n \in \Iset_{\Cset}}  E(n) k_n \nonumber \\
\text{subject to} \mbox{   } &  \max_{n \in \Iset_{\Cset}} \{ \Delta(n) k_n \} \leq \Delta_{\text{thr}} \nonumber \\
&  \sum_{n \in \Iset_{\Cset}} k_n = K, 
\end{align} where the delay constraint is $\max_{n \in \Iset_{\Cset}} \{ \Delta(n) k_n \} \leq \Delta_{\text{thr}}$ instead of $\sum_{n \in \Iset_{\Cset}} \Delta(n) k_n \leq \Delta_{\text{thr}}$ in (\ref{eq:serial_opt_problem}) thanks to parallel processing. 

The optimal solution to (\ref{eq:parallel_opt_problem}) orders devices depending on their average energy consumption $E(n)$ (in increasing order). The vector of ordered devices is $\boldsymbol d_e$, where $[\boldsymbol d_e]_r$ is the $r$th element of the vector $\boldsymbol d_e$. The optimal solution assigns $k_{[\boldsymbol d_e]_1} = \floor*{\frac{\Delta_{\text{thr}}}{\Delta({[\boldsymbol d_e]_1})}}$ to device $[\boldsymbol d_e]_1$. If there still exist tasks waiting to be scheduled, it continues assigning tasks to $[\boldsymbol d_e]_2, \ldots, [\boldsymbol d_e]_{N+1}$ one by one using the same rule and stops when all the tasks are scheduled as summarized in Algorithm~\ref{alg:parallel}. 

\begin{algorithm}
\caption{The optimal solution for parallel task allocation}\label{alg:parallel}
\begin{algorithmic}[1]
\State $K_{\text{sch}} = K$. $r = 1$. $k_{[\boldsymbol d_e]_r} = 0$, $\forall r \in \{1, \ldots, N+1\}$.
\While{$K_{\text{sch}} > 0 $ AND $r \leq N+1$} 
\State Assign $k_{[\boldsymbol d_e]_r} = \floor*{\frac{\Delta_{\text{thr}}}{\Delta({[\boldsymbol d_e]_r})}}$ tasks to device $[\boldsymbol d_e]_r$ 
\State $K_{\text{sch}} = \max\{0, K_{\text{sch}} - k_{[\boldsymbol d_e]_r}\}$. $r = r + 1$
\EndWhile
\end{algorithmic}
\end{algorithm}

Our online algorithm mimics the offline solution in Algorithm~\ref{alg:parallel}. At the start (when scheduling starts), our algorithm runs Algorithm~\ref{alg:parallel}. If $k_{[\boldsymbol d_e]_r} > 0$, one task is assigned to device $[\boldsymbol d_e]_r$. Then, periodically or when a device finishes processing a task, Algorithm~\ref{alg:parallel} is run again and if $k_{[\boldsymbol d_e]_r} > 0$, a task is assigned to device $[\boldsymbol d_e]_r$. This procedure continues until all tasks are successfully scheduled or hard deadline constraint is reached. As compared to Algorithm~\ref{alg:parallel}, our online algorithm assigns tasks to devices one by one, which better adapts to the time-varying  resources at edge devices.

\section{\label{sec:results} Performance Evaluation}
In this section, we evaluate the performance of our algorithm; \MaCC \ for serial and parallel setups using Android-based smartphones. %\footnote{We note that extensive experiment results including different mobility models, multiple master / worker scenarios are omitted in this paper due to page limitations, but provided in our technical report \cite{MaCC_tech_rep}.} 
We implemented a testbed of a master and multiple workers using real mobile devices, specifically Android 6.0.1 based Nexus 6P and Nexus 5 smartphones. Nexus 6P has higher energy efficiency than Nexus 5. All the workers are connected to the master device using Wi-Fi Direct connections. We conducted our experiments using our testbed in a lab environment where several other Wi-Fi networks were operating in the background. We located all the devices in close proximity of each other  (within a few meters distance). 

\begin{figure}[t!]
	\centering
	\vspace{-10pt}
	\subfigure[Task completion time]{ {\includegraphics[height=33mm]{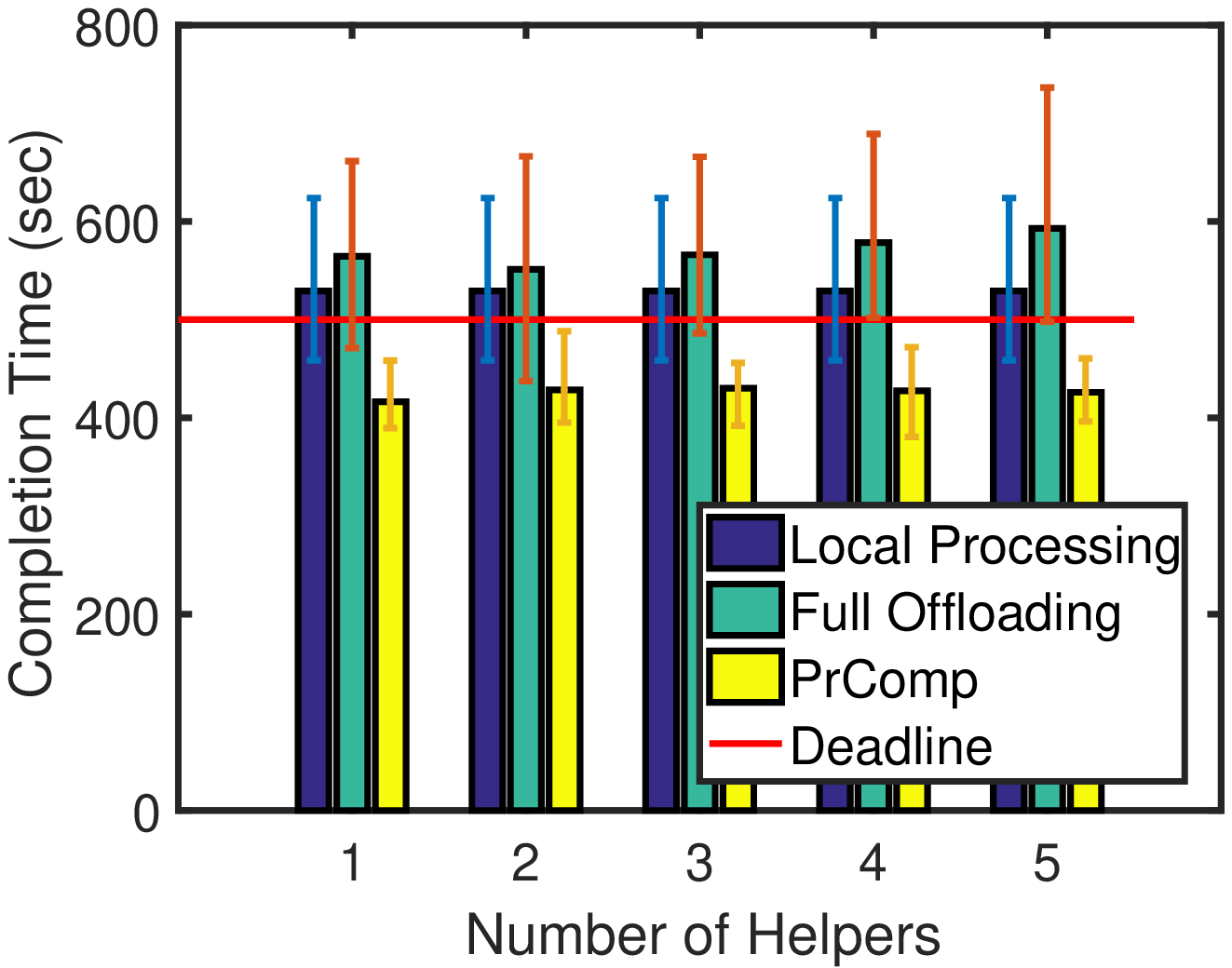}} } \hspace{-20pt}
	\subfigure[Energy consumption]{ {\includegraphics[height=32mm]{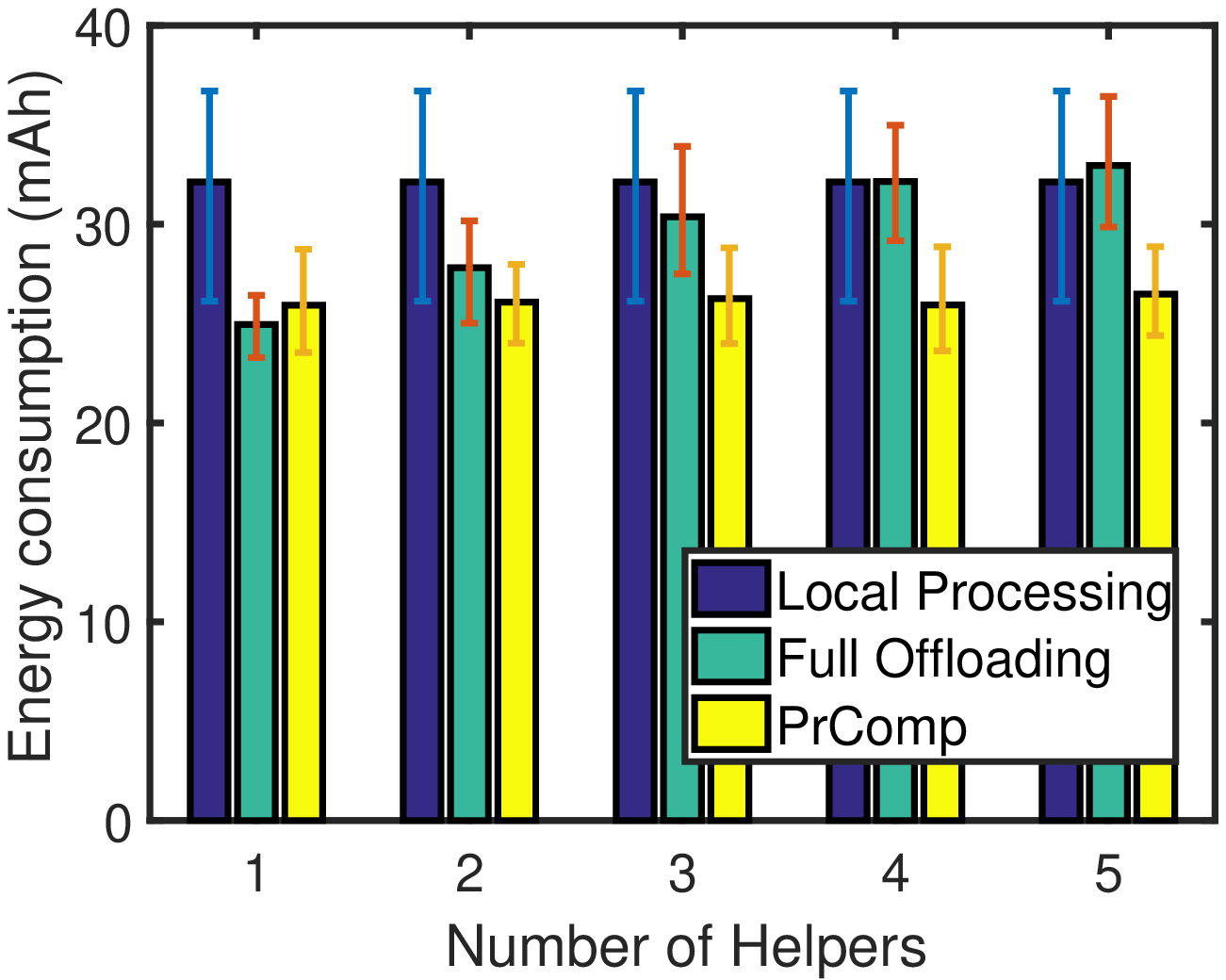}} }
	\vspace{-5pt}
	\caption{The task completion time and energy consumption performance of \MaCC \ for serial tasks. The figures are generated by averaging 16 trials. 60 images are processed and the deadline threshold is 500 sec. 
	}
	\vspace{-10pt}
	\label{fig:results_serial}
\end{figure}

Fig.~\ref{fig:results_serial} shows the performance of \MaCC \ for serial tasks, where we used the face detection application, similar to the setup in Section~\ref{sec:res_prediction}, as a serial task. The master device is Nexus 5, and the workers are Nexus 6P. The performance of \MaCC \ is evaluated as compared to baselines: (i) {\tt Full Offloading}, which offloads each task to a worker device that has the least energy consumption, but does not allow local processing. This baseline is similar to the algorithm developed in \cite{ARC}, but updated for serial tasks setup, (ii) {\tt Local Processing}, where the master device processes all the tasks (\ie it does not offload tasks). We assume {\em statistical} mobility model described in Section~\ref{sec:res_prediction}, %(other mobility models and their corresponding experiments are provided in \cite{MaCC_tech_rep}), 
where time is divided into 10 sec slots. At each slot, one of the helpers moves out of transmission range of the master device with probability 0.3, and comes back to the transmission range with probability 0.5. For the other helpers, these (both moving out and in) probabilities are 0.9. Given these values, it is straightforward to calculate $P_{\pi_{k,l}}$. % as detailed in our technical report \cite{MaCC_tech_rep}. 
%
%become in the same transmission range with the helper with probability 0.3, and 
%
%The time step for mobility is 10 seconds, with one helper 0.3/0.5 and the rest 0.9/0.9. }
%
Fig.~\ref{fig:results_serial} (a) shows the task completion time versus number of helpers, and Fig.~\ref{fig:results_serial} (b) shows the total energy consumption (at all masters and helpers). As seen \MaCC \ satisfies the hard deadline constraint and significantly reduces task  completion time,  while {\tt Local Processing} fails to satisfy the deadline constraint on the average and {\tt Full Offloading} fails to satisfy the deadline constraint in some instances (confidence interval exceeds deadline). {\tt Full Offloading} performs worse in terms of both delay and energy consumption, because the master device is Nexus 5 which is a weaker device as compared to workers (Nexus 6P). The energy consumption of {\tt Full Offloading} increases with increasing number of helpers, because more workers cause more energy consumption. \MaCC \ performs the best as it (i) takes advantage of local resources at the master device as well as workers, and (ii) is adaptive to time-varying resources.

\begin{figure}[t!]
	\centering
	%\vspace{-10pt}
	%\includegraphics[height=40mm]{Fig_Yuxuan/basic_time_new.eps}
	\subfigure[Task completion time]{ {\includegraphics[height=33mm]{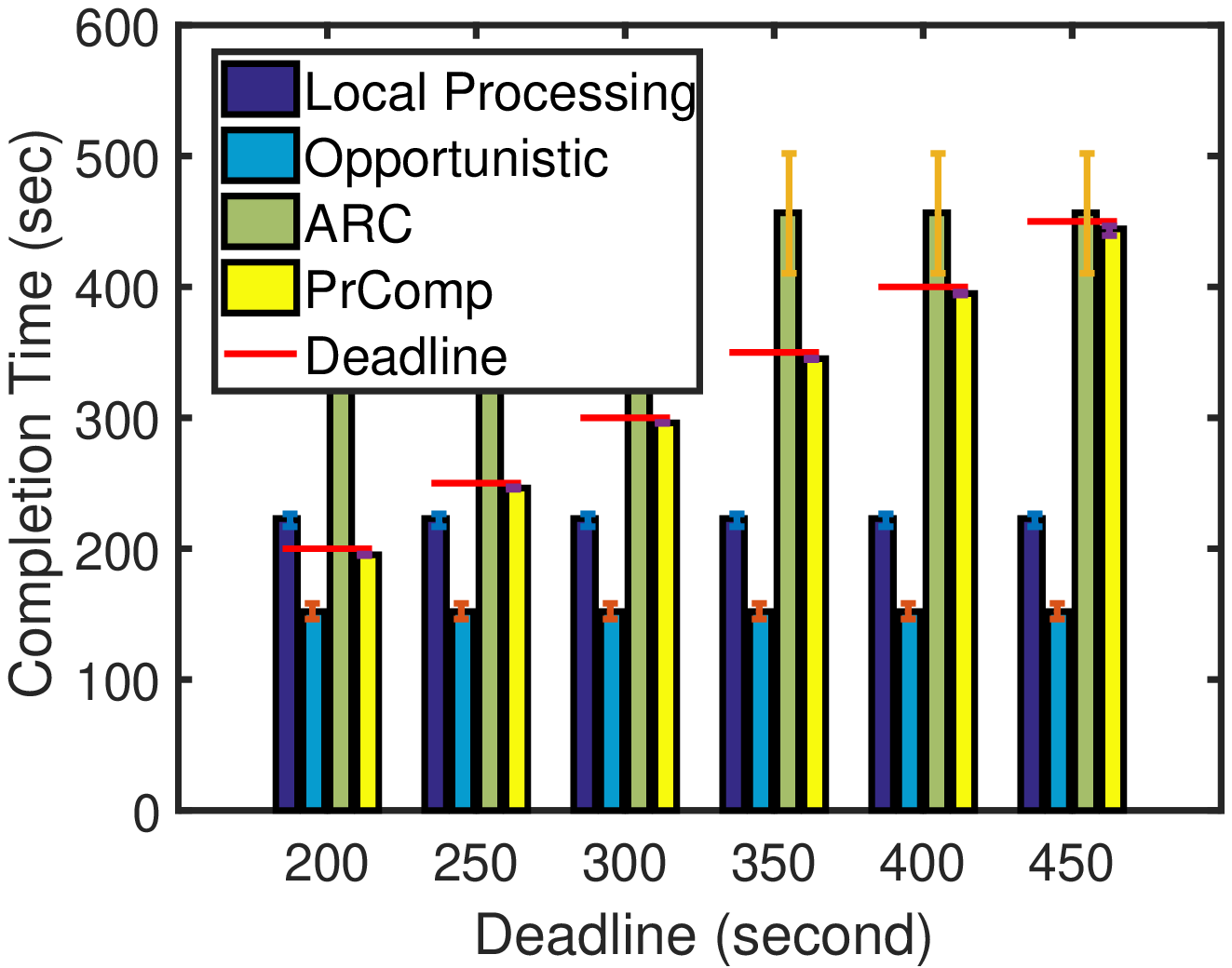}} } \hspace{-20pt}
	\subfigure[Energy consumption]{ {\includegraphics[height=32mm]{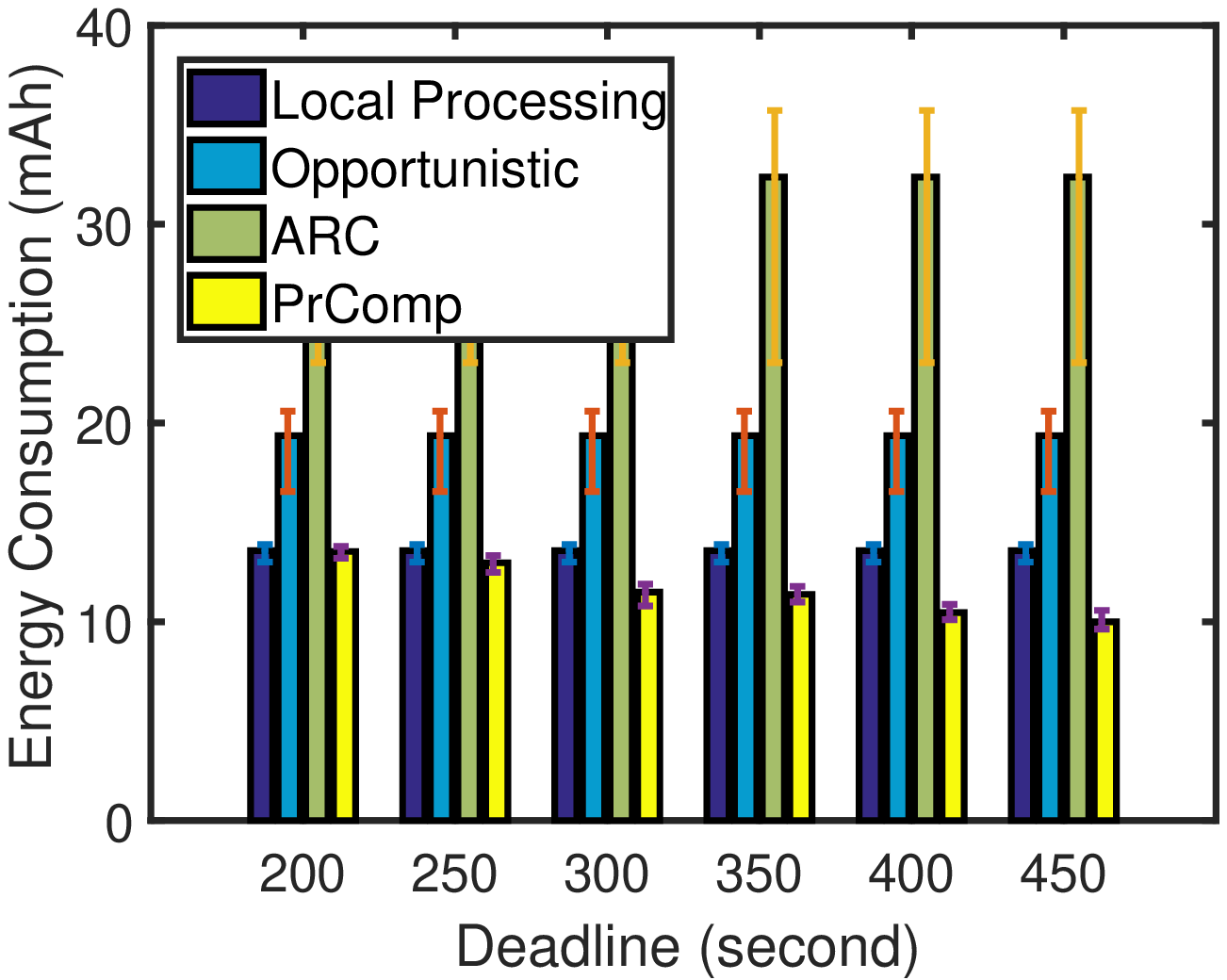}} }
	\vspace{-5pt}
	\caption{The task completion time and energy consumption performance of \MaCC \ for parallel tasks. The figures are generated by averaging 16 trials. 
		%
		%Completion time and energy consumption for local processing, Opportunistic and MaCC Para when there are 2 helpers with task-based mobility (0.1 and 0.8)
	}
	\vspace{-10pt}
	\label{fig:results_parallel}
\end{figure}

Fig.~\ref{fig:results_parallel} shows the delay and energy performance of \MaCC \ for parallel tasks, where we used matrix multiplication $Y=AX$ as a parallel task. $A$ is a $10\text{K} \times 10\text{K}$ matrix, $X$ is a $10\text{K} \times 1$ vector. Matrix $A$ is divided into 500 sub-matrices, each of which is a $20 \times 10\text{K}$ matrix. There is a master device (Nexus 5) and two workers (Nexus 6P). The probability of not being in the same transmission range of workers are $P_1 = 0.1$ and $P_2 = 0.8$.  \MaCC \ is compared with baselines: {\tt Local Processing}, which is the same algorithm described above; {\tt Opportunistic}, which uses master and worker devices simultaneously; and {\tt ARC}, which is an algorithm developed in \cite{ARC} to reduce the energy consumption at local devices (\ie the master device). As seen in Fig.~\ref{fig:results_parallel}(a), \MaCC, although it has larger task completion time as compared to baselines, it always satisfies  hard deadline constraints. Furthermore, \MaCC \ reduces total energy consumption as compared to baselines, and its energy efficiency increases when the hard deadline threshold increases, because \MaCC \ has a larger set of task scheduling policies that it can exploit when deadline threshold increases. 

\begin{figure}[t!]
	\centering
	%\vspace{-10pt}
	%\includegraphics[height=40mm]{Fig_Yuxuan/basic_time_new.eps}
	\subfigure[Task completion time]{ {\includegraphics[height=33mm]{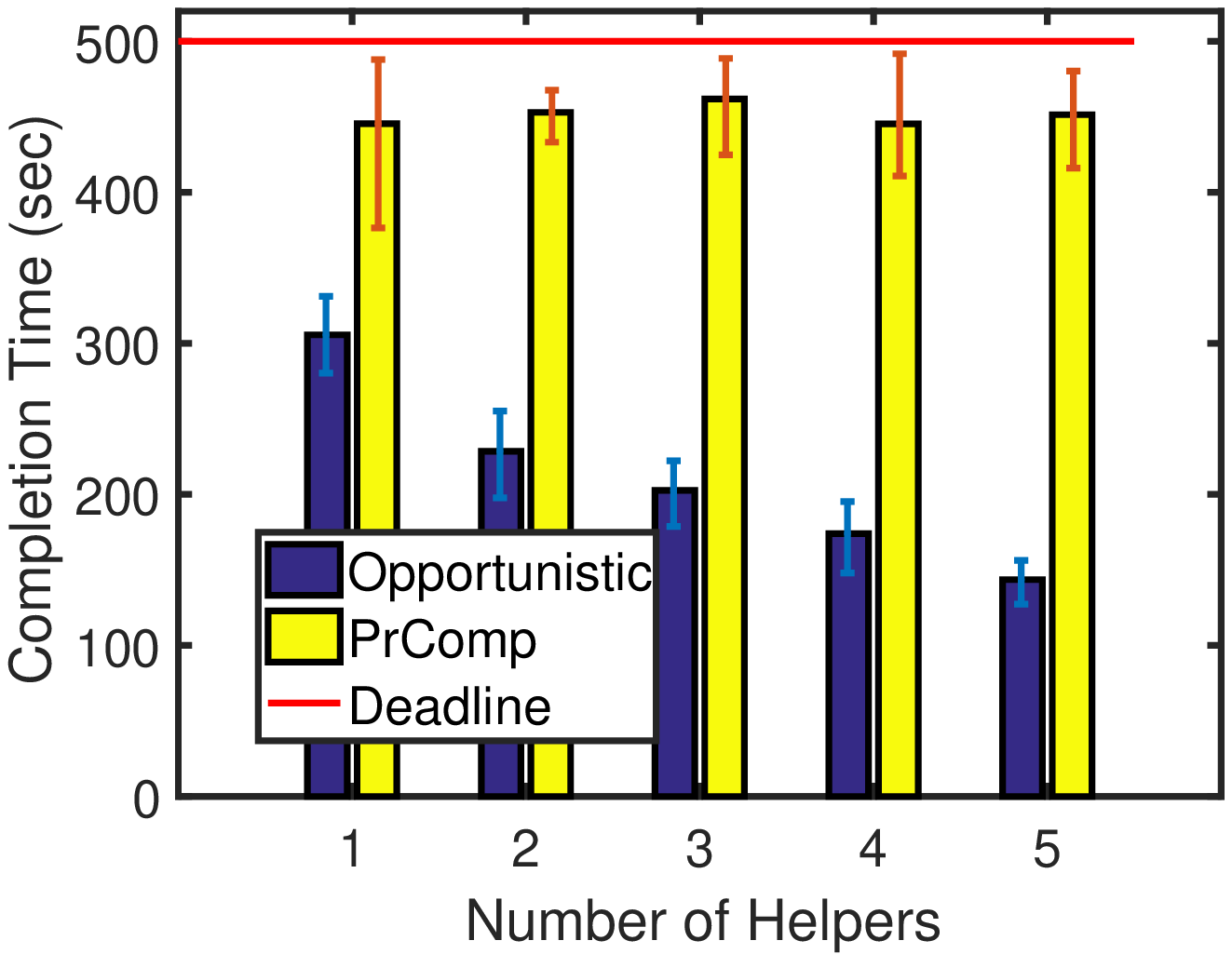}} } \hspace{-20pt}
	\subfigure[Energy consumption]{ {\includegraphics[height=32mm]{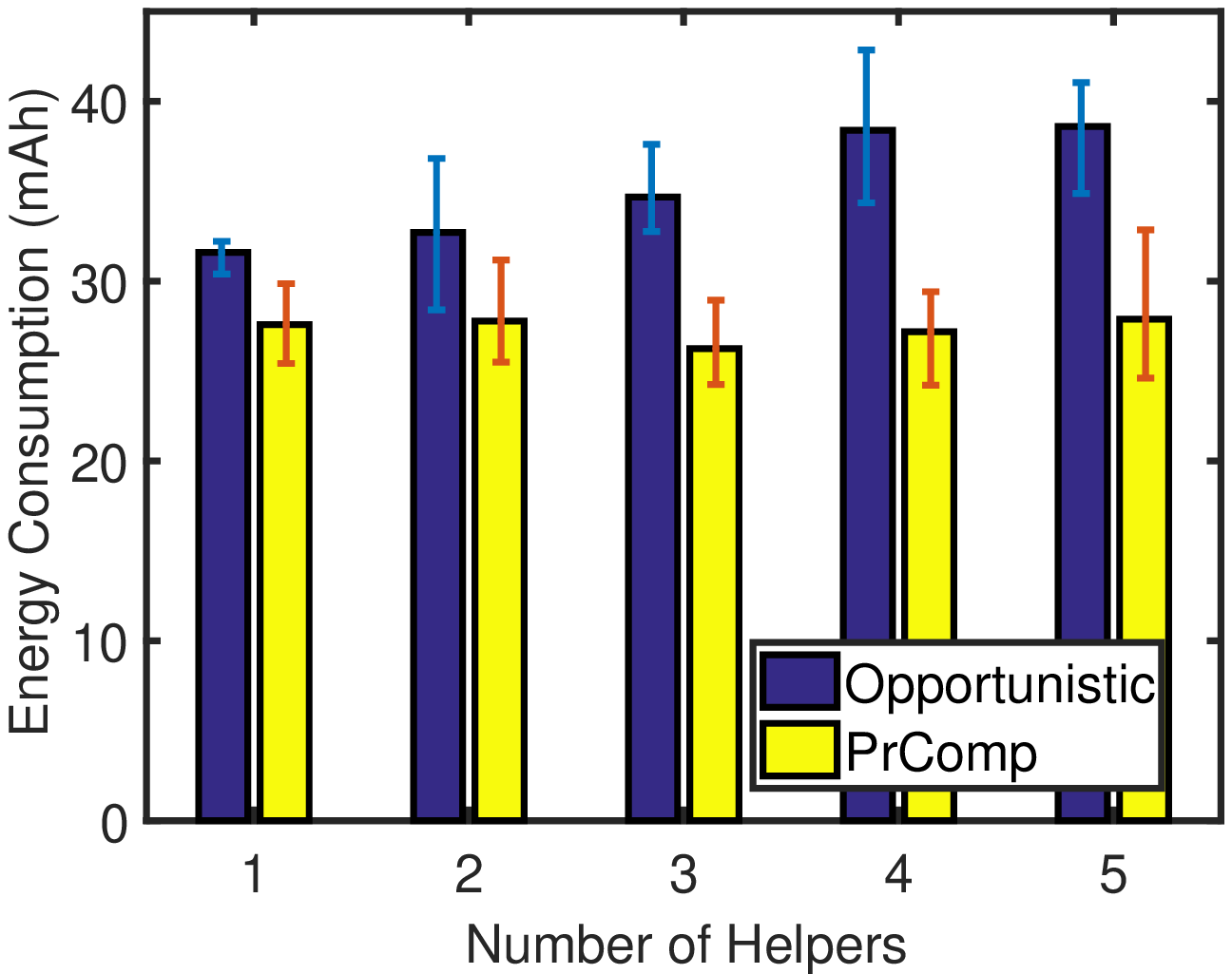}} }
	\vspace{-5pt}
	\caption{The task completion time and energy consumption performance of \MaCC \ versus number of helpers for parallel tasks (face detection). The figures are generated by averaging 5 trials. The probability of not being in the same transmission range of workers are $P_1 = 0.3$ and $P_2 = \ldots = P_5 = 0.6$.
	}
	\vspace{-10pt}
	\label{fig:results_parallel_face_06}
\end{figure}

\begin{figure}[t!]
	\centering
	%\vspace{-10pt}
	%\includegraphics[height=40mm]{Fig_Yuxuan/basic_time_new.eps}
	\subfigure[Task completion time]{ {\includegraphics[height=33mm]{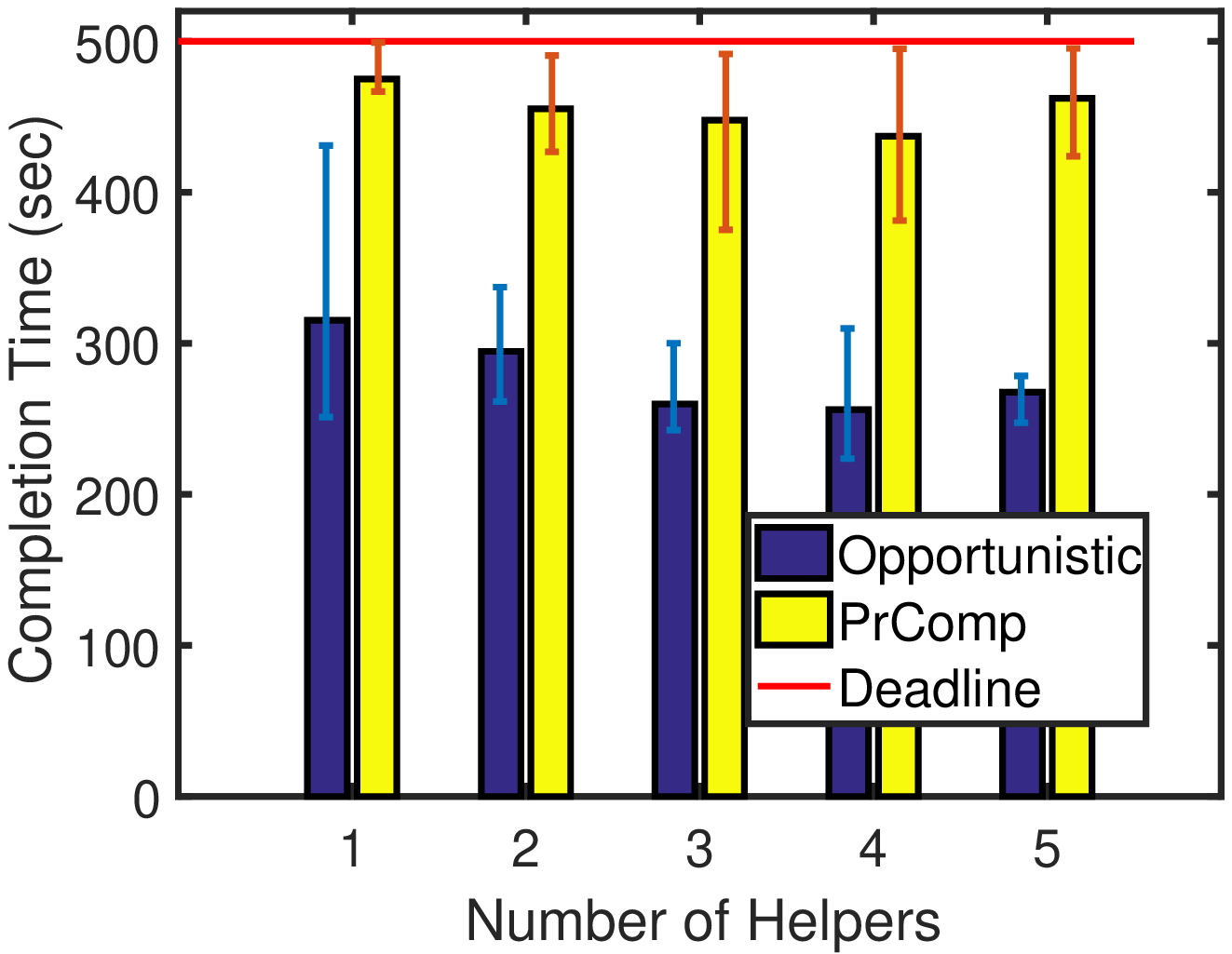}} } \hspace{-20pt}
	\subfigure[Energy consumption]{ {\includegraphics[height=32mm]{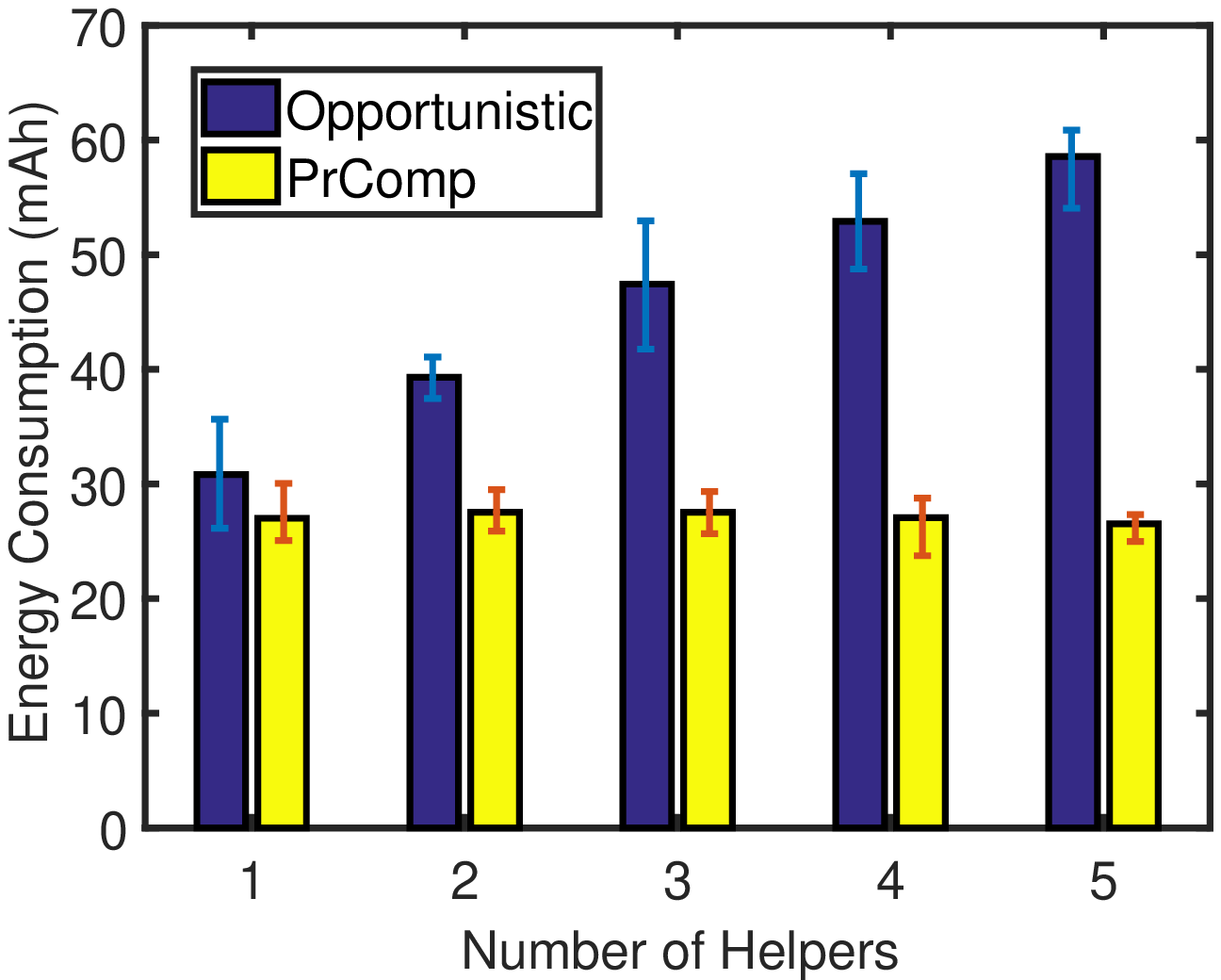}} }
	\vspace{-5pt}
	\caption{The task completion time and energy consumption performance of \MaCC \ versus number of helpers for parallel tasks (face detection). The figures are generated by averaging 5 trials. The probability of not being in the same transmission range of workers are $P_1 = 0.3$ and $P_2 = \ldots \P_5 = 0.9$. 
	}
	\vspace{-10pt}
	\label{fig:results_parallel_face}
\end{figure}

Fig.~\ref{fig:results_parallel_face_06} and Fig.~\ref{fig:results_parallel_face} demonstrate the delay and energy performance of \MaCC \ versus the number of helpers for the same setup described above, but we use face detection as a parallel task. The hard deadline is 500 seconds. The probability of not being in the same transmission range of workers are $P_1 = 0.3$ and $P_2 = \ldots = P_5 = 0.6$ and $P_1 = 0.3$ and $P_2 = \ldots P_5 = 0.9$ for Fig.~\ref{fig:results_parallel_face_06} and Fig.~\ref{fig:results_parallel_face}, respectively. As seen, \MaCC \ always satisfies the hard deadline constraints and performs better in terms of energy consumption when mobility of workers increases thanks to making task offloading decisions by taking into account the mobility of devices. 

\begin{figure}[t!]
	\centering
	%\vspace{-10pt}
	%\includegraphics[height=40mm]{Fig_Yuxuan/basic_time_new.eps}
	\subfigure[Task completion time]{ {\includegraphics[height=33mm]{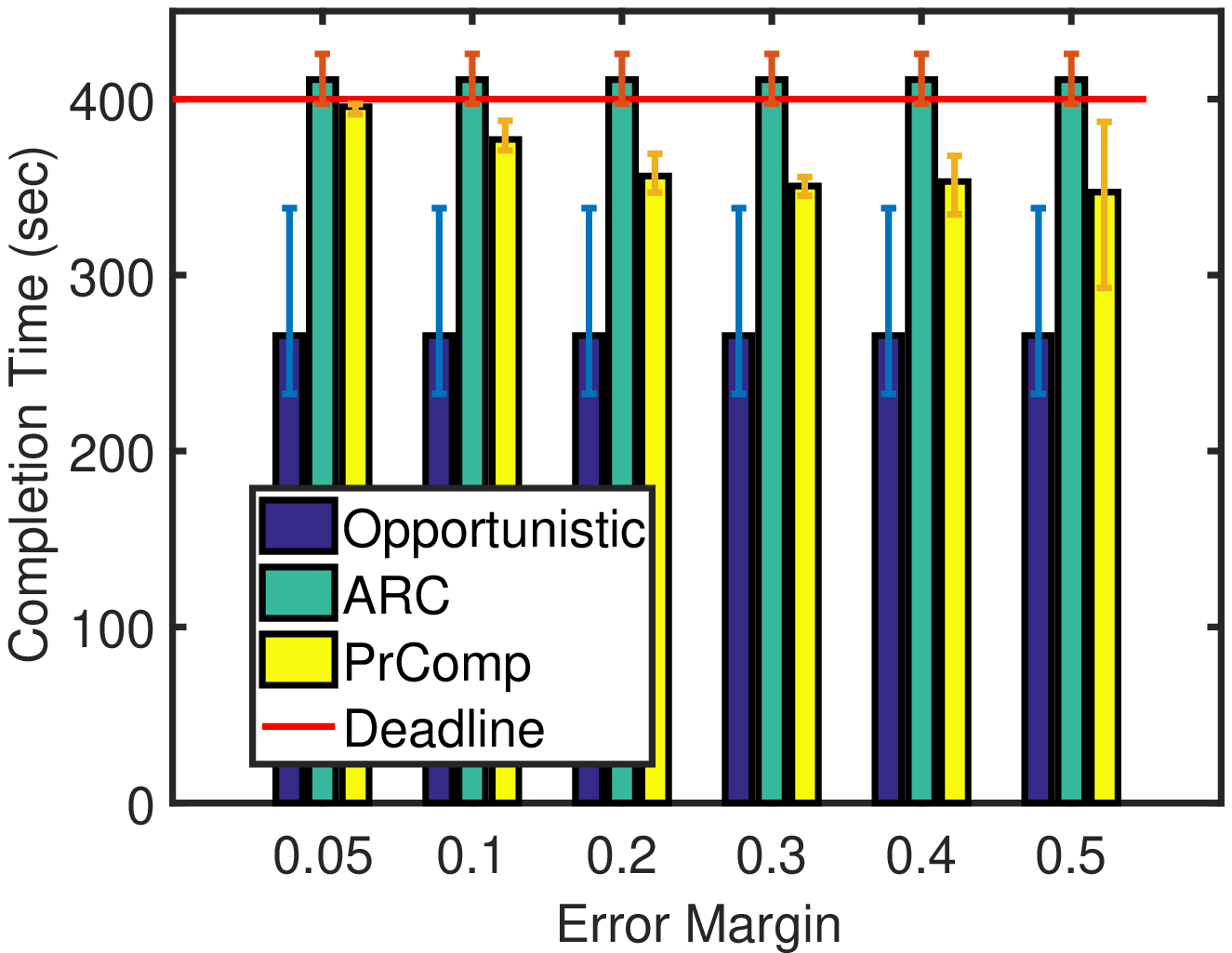}} } \hspace{-20pt}
	\subfigure[Energy consumption]{ {\includegraphics[height=32mm]{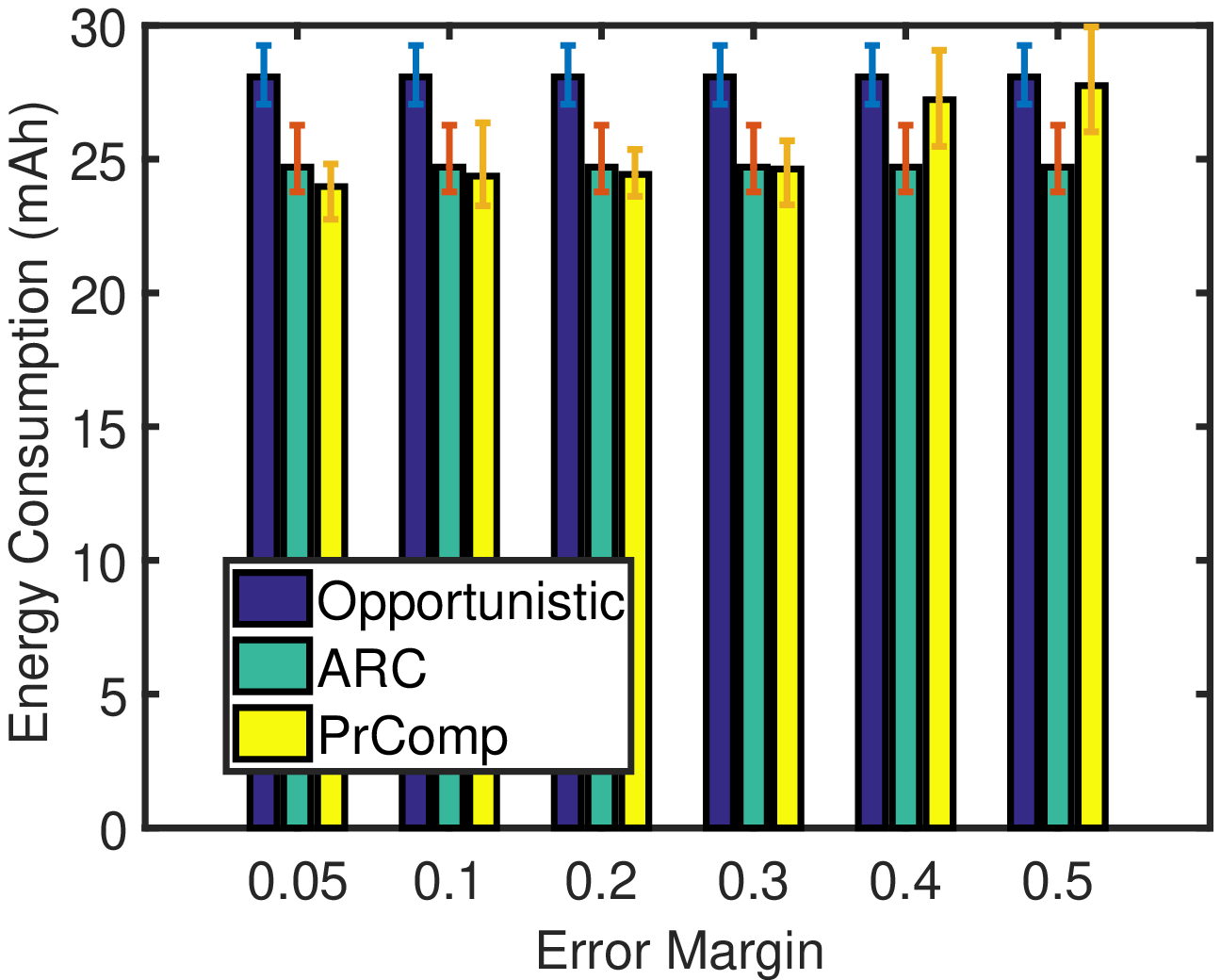}} }
	\vspace{-5pt}
	\caption{The task completion time and energy consumption performance of \MaCC \ versus error margin for parallel tasks (face detecton) when workers follow real mobility data in \cite{dataset}. The figures are generated by averaging 5 trials. 
	}
	\vspace{-10pt}
	\label{fig:match}
\end{figure}

Fig.~\ref{fig:match} shows the task completion time and energy consumption performance of \MaCC \ versus error margin. % for parallel tasks (face detection) when workers follow real mobility patterns collected in \cite{dataset}. 
Master device is Nexus 5 and we have three worker devices; all of them are Nexus 6P. The hard deadline constraint is 400 seconds. The worker devices follow mobility pattern from the dateset in \cite{dataset}, which collects data on if the master and a worker device is in the same transmission range or not. The master device estimates whether a worker device is in its transmission range with some error probability, which corresponds to the error margin. 

\MaCC \ is compared with baselines: {\tt Opportunistic} and {\tt ARC}. As seen in Fig.~\ref{fig:match} (a), both {\tt Opportunistic} and \MaCC \ satisfies the hard deadline while the average completion time of {\tt ARC} exceeds the deadline. The energy consumption of \MaCC \ is less than {\tt Opportunistic} as seen in Fig.~\ref{fig:match} (b), but higher than {\tt ARC} as {\tt ARC} is optimized for energy, but {\tt ARC} misses the hard deadline constraint as seen in Fig.~\ref{fig:match} (a). 
%The completion time of \MaCC \ has a decreasing trend when the error margin increases. This is because error margin is treated as the probability of out of range and when the error margin is higher, \MaCC \ attempts to offload more subtasks simultaneously to meet the deadline. On the aspect of energy consumption shown in Fig.~\ref{fig:match} (b), when the error margin is 0.4 and above, \MaCC \ consumes more energy than {\tt ARC}. Under these circumstances, master will assume local processing is the most energy efficient solution. Hence it will choose local processing many more times, which leads to higher energy consumption.  

Table.~\ref{table:majority} shows the delay and energy performance of \MaCC \ when majority voting is used to predict the mobility of devices. These results are for the parallel task setup, where we used the face detection application as a parallel task. The master device is Nexus 5 and workers are Nexus 6P smartphones. There are three workers. All workers move according to the mobility pattern from the dataset in \cite{dataset}. In order to predict the mobility, we divide the time into slots as described in Section~\ref{sec:mobility}, and the master device counts the number of encounters in the last 11 slots with each worker device. If there are encounters during most of the slots with a worker, the master concludes that it will encounter with this worker at the next time slot. Table~\ref{table:majority} shows that \MaCC \ significantly improves task completion time as well as energy consumption as compared to {\tt Local Processing} thanks to effectively using resources at master and workers by predicting mobility, while  {\tt Local Processing} is limited with the resources at the master device. 

\begin{table}[h!]
	\caption{Completion time and energy consumption performance of \MaCC \ when mobility is predicted via majority voting.}
	\centering
	\begin{tabular}{|c|c|c|}
		\hline
		& \MaCC & {\tt Local Processing} \tabularnewline
		%\hline
		\hline
		Completion Time (sec) & 454.4 & 529.1  \tabularnewline
		\hline
		Energy Consumption (mAh)  & 28.82 &  32.12 \tabularnewline
		%\hline
		%$N=3$  & $4$ & $4$\tabularnewline
		%\hline
		%$N=4$  & $3$ & $3$\tabularnewline
		\hline
	\end{tabular}
	\label{table:majority}
\end{table}

\begin{figure}[t!]
	\centering
	%\vspace{-10pt}
	%\includegraphics[height=40mm]{Fig_Yuxuan/basic_time_new.eps}
	\subfigure[Trial 1]{ {\includegraphics[height=33mm]{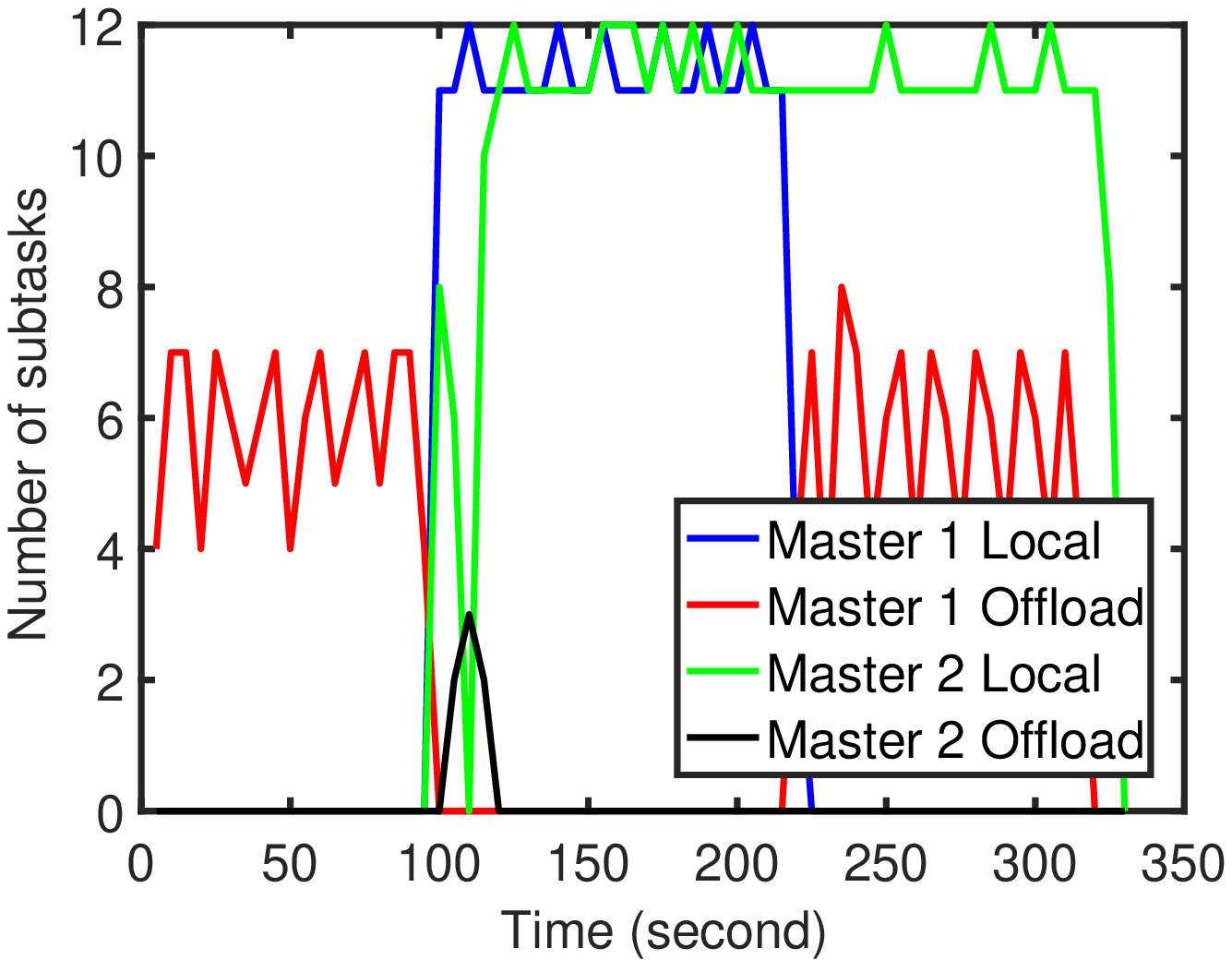}} } \hspace{-20pt}
	\subfigure[Trial 2]{ {\includegraphics[height=32mm]{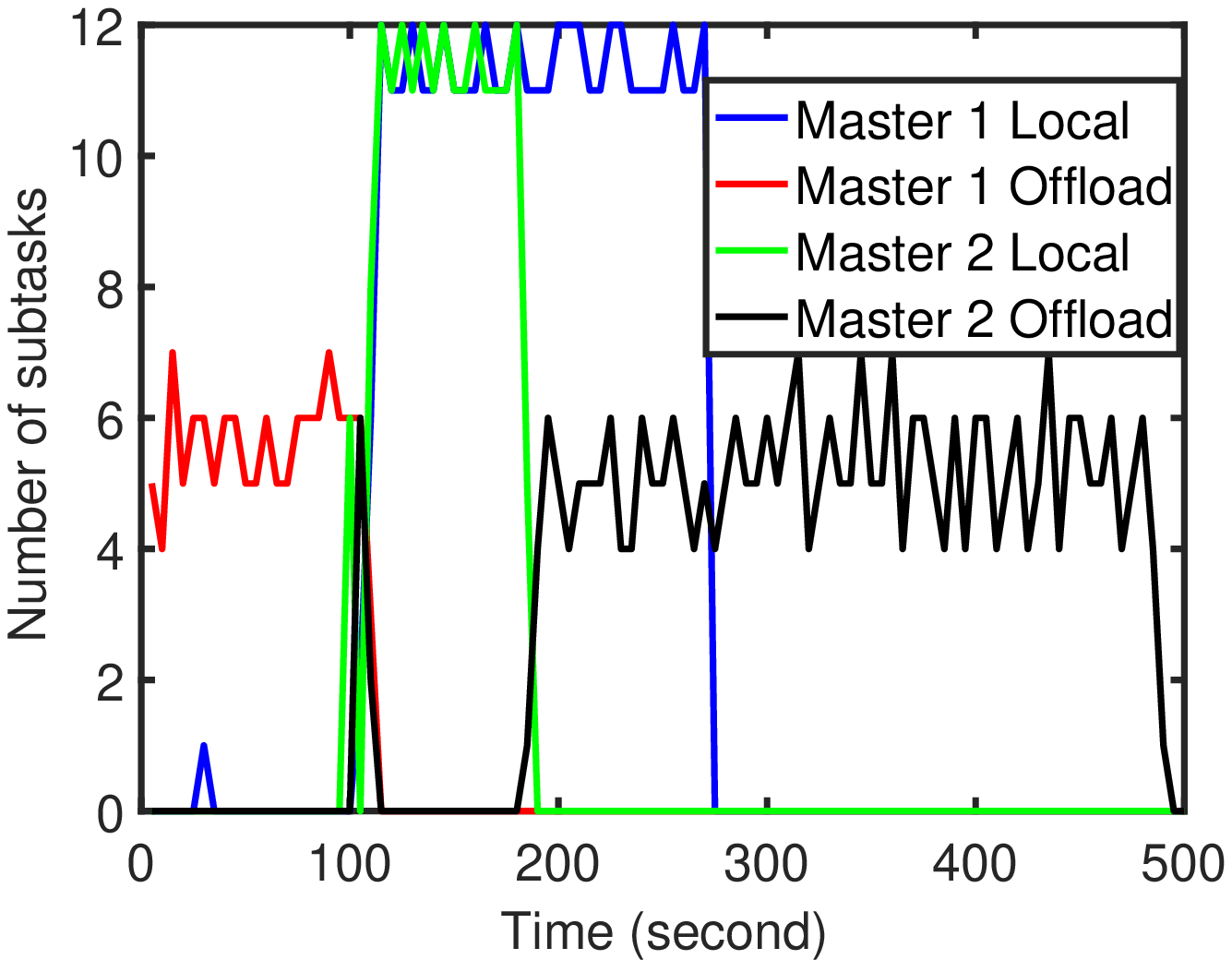}} }
	\vspace{-5pt}
	\caption{The number of subtasks at each device versus time for the parallel task setup. There are two masters and one worker. 
	}
	\vspace{-10pt}
	\label{fig:multi_masters}
\end{figure}

Now, we focus on the scenario of multiple master devices. In particular, we have two master devices connected to one worker device. The worker device serves to both master devices simultaneously by multi-threading. We use matrix multiplication ($Y=AX$) as a parallel task in this setup. $A$ is a $10\text{K} \times 10\text{K}$ matrix, $X$ is a $10\text{K} \times 1$ vector. Matrix $A$ is divided into 500 sub-matrices, each of which is a $20 \times 10\text{K}$ matrix. Both masters are Nexus 5 smartphones, while the worker is a Nexus 6P smartphone. The hard deadline constraint is 500 seconds. All devices are always in the same transmission range in this scenario (\ie there is no mobility). The task of master 1 starts at time 0, while the task of master 2 starts after 100 seconds. Fig.~\ref{fig:multi_masters} shows two trials of the number of processed sub-tasks at master and worker devices versus time. As seen, Master 1 always offloads its tasks to the worker device as it is more energy efficient (as master is Nexus 5 and worker is Nexus 6P) before reaching 100 second threshold. After this point, both Master 1 and 2 try to offload their tasks to the worker, but they immediately realize that the worker device becomes less efficient (in terms of both delay and energy consumption). Thus, master devices back off and process their tasks locally. After both masters process their tasks locally for a while, they will probe the worker device. In Fig.~\ref{fig:multi_masters} (a), Master 1 starts offloading to the worker first, while in Fig.~\ref{fig:multi_masters} (b), Master 2 starts offloading to the worker earlier than Master 2. This figure shows that our algorithm works in the multiple worker scenario, where the worker resources (especially energy) is utilized efficiently, \ie multiple masters does not drain all the energy in the worker device. Fig.~\ref{fig:multi_masters} also shows that our algorithm has room for improvement to provide fair share of worker resources across multiple master devices.

%\input{MACC}
%\input{evaluation}
%\vspace{-5pt}
\section{\label{sec:conclusion}Conclusion}
%\vspace{-5pt}
We developed a predictive edge computing algorithms \MaCC \ with hard deadline constraints for serial and parallel tasks. Our algorithms (i) predict the uncertain dynamics of resources of edge devices, and (ii) make task offloading decisions by taking into account the predicted available resources, as well as the hard deadline constraints of tasks. We evaluate \MaCC \ on a testbed consisting of real Android-based smartphones. The experiments  show that \MaCC \ algorithms significantly improve energy consumption of edge devices as well as task completion delay as compared to baselines.

\bibliographystyle{IEEEtran}
%\bibliography{IEEEabrv,refs}
\bibliography{refs}

%\vspace{-2.5mm}
%\bibliographystyle{IEEEtran}
%\bibliography{src/bib}

\end{document}